\documentclass[epj]{svjour}

\usepackage{amsmath}
\usepackage{graphicx}

\begin{document}

\title{Loop-corrected belief propagation for lattice spin models
\thanks{The final publication is available at Springer via 
{\tt http://dx.doi.org/10.1140/epjb/e2015-60485-6}}}

\author{Hai-Jun Zhou \thanks{Email address: {\tt zhouhj@itp.ac.cn}.}
  and Wei-Mou Zheng}

\institute{State Key Laboratory of Theoretical Physics, Institute of
  Theoretical Physics, Chinese Academy of Sciences, Beijing 100190, China}

\date{18 June 2015; revised 28 August 2015; further revised 20 October 2015}

\abstract{
  Belief propagation (BP) is a message-passing method for solving probabilistic
  graphical models. It is very successful in treating disordered models (such
  as spin glasses) on random graphs. On the other hand, finite-dimensional
  lattice models have an abundant number of short loops, and the BP method is
  still far from being satisfactory in treating the complicated loop-induced
  correlations in these systems. Here we propose a loop-corrected BP method to
  take into account the effect of short loops in lattice spin models. We
  demonstrate, through an application to the square-lattice Ising model, that
  loop-corrected BP improves over the naive BP method significantly. We also
  implement loop-corrected BP at the coarse-grained region graph level to
  further boost its performance.
  \PACS{
    {02.70.Rr}{General statistical methods}  \and
    {75.10.Nr}{Spin-glass and other random models} \and
    {07.05.Pj}{Image processing} \and
    {05.50.+q}{Lattice theory and statistics (Ising, Potts, etc.)}
  }
}

\maketitle

\section{Introduction}

Belief propagation (BP) is a message-passing method for solving probabilistic
graphical models. It was developed in the computer science research field
\cite{Pearl-1988} and, independently, also in the statistical physics field
along with the replica-symmetric mean field theory \cite{Mezard-etal-1987}.
For spin glass physicists the BP method is commonly referred to as the 
replica-symmetric cavity method.
The basic physical idea behind BP is the Bethe-Peierls approximation
\cite{Bethe-1935,Peierls-1936a,Chang-1937}, which assumes that if a vertex is
deleted from a graph, all of its nearest neighboring vertices will become
completely uncorrelated in the remaining (cavity) graph. BP has good
quantitative predicting power if the graph's characteristic loop length is much
longer than the system's typical correlation length.

The BP method is exact for models defined on a tree graph which contains no
loops. A finite-connectivity random graph contains many loops, but the
typical loop length increases logarithmically with the total number of vertices
in the graph, and BP also performs excellently on sufficiently large
random-graph systems. A lot of random combinatorial optimization problems and
random-graph spin glass models have been successfully solved by BP and the
replica-symmetric mean field theory during the last two decades
\cite{Mezard-Montanari-2009}.

Finite-dimensional lattice models have an abundant number of short loops, which
cause complicated local correlations in the system. The correlation length of
the system at sufficiently low temperatures often exceeds the characteristic
length of short loops. At the moment, BP is still far from being
satisfactory in treating the complicated loop-induced local correlations in
these systems. In recent years the generalized belief propagation (GBP),
as a promising way of overcoming the naive BP's shortcomings, has been
seriously explored 
\cite{Yedidia-Freeman-Weiss-2005,Pelizzola-2005,Rizzo-etal-2010,Wang-Zhou-2013,LageCastellanos-Mulet-RicciTersenghi-2014}.
The GBP method is rooted in the cluster variational method
\cite{Kikuchi-1951,An-1988} and it abandons the Bethe-Peierls approximation.

Here we explore a simple way of improving  BP while
still keeping the Bethe-Peierls approximation. We propose a loop-corrected BP
method to take into account the effect of short loops in lattice spin models.
The loop-corrected BP method, as a hierarchical approximation scheme, is
conceptually straightforward to understand, and its numerical implementation
appears to be easier than the GBP method. As a proof of principle, we apply
loop-corrected BP to the square-lattice Ising model for which exact results are
available, and demonstrate that it indeed significantly outperforms the naive
BP. We also implement loop-corrected BP at the coarse-grained region graph level
\cite{Zhou-Wang-2012} to further boost its performance. Our numerical results
on the square-lattice Ising model indicate that loop-corrected BP might be a
preferred method than GBP.

The actual applications of loop-corrected BP to the Edwards-Anderson spin
glass model \cite{Edwards-Anderson-1975} on the square lattice and especially
on the three-dimensional cubic lattice will be carried out in a follow-up paper.
As potential practical applications, we suggest that loop-corrected BP might be
useful in two-dimensional image processing tasks, such as image recovery
\cite{Nishimori-2001}.

For reason of clarity, in the remaining part of this paper we describe the
loop-corrected BP method using the square lattice as a representative example. 

Let us finish the Introduction by noting that loop correction to BP
has been a focusing issue in the last decade and various
protocols have been investigated
\cite{Montanari-Rizzo-2005,Parisi-Slanina-2006,Chertkov-Chernyak-2006b,Mooij-etal-2007,Bulatov-2008,Xiao-Zhou-2011,Zhou-Wang-2012,Ravanbakhsh-Yu-Greiner-2012}). For example, the proposal of Mooij and co-authors
\cite{Mooij-etal-2007} (see also \cite{Ravanbakhsh-Yu-Greiner-2012})
considers the correlations
among the neighboring vertices of a given focal vertex by computing the
joint distribution of all these neighboring vertices' spin states. This
proposal abandons the Bethe-Peierls approximation and it is computationally
expensive, while its performance on square-lattice spin models seems 
to be inferior to that of the conventional cluster variational method 
\cite{Mooij-etal-2007}. Another interesting approach \cite{Bulatov-2008} is
based on the self-avoiding walk tree representation of a loopy graph
\cite{Weitz-2006} and its full potential is yet to be explored.

\section{The lattice spin system}
\label{sec:system}

\begin{figure}
  \begin{center}
    \includegraphics[width=0.3\textwidth]{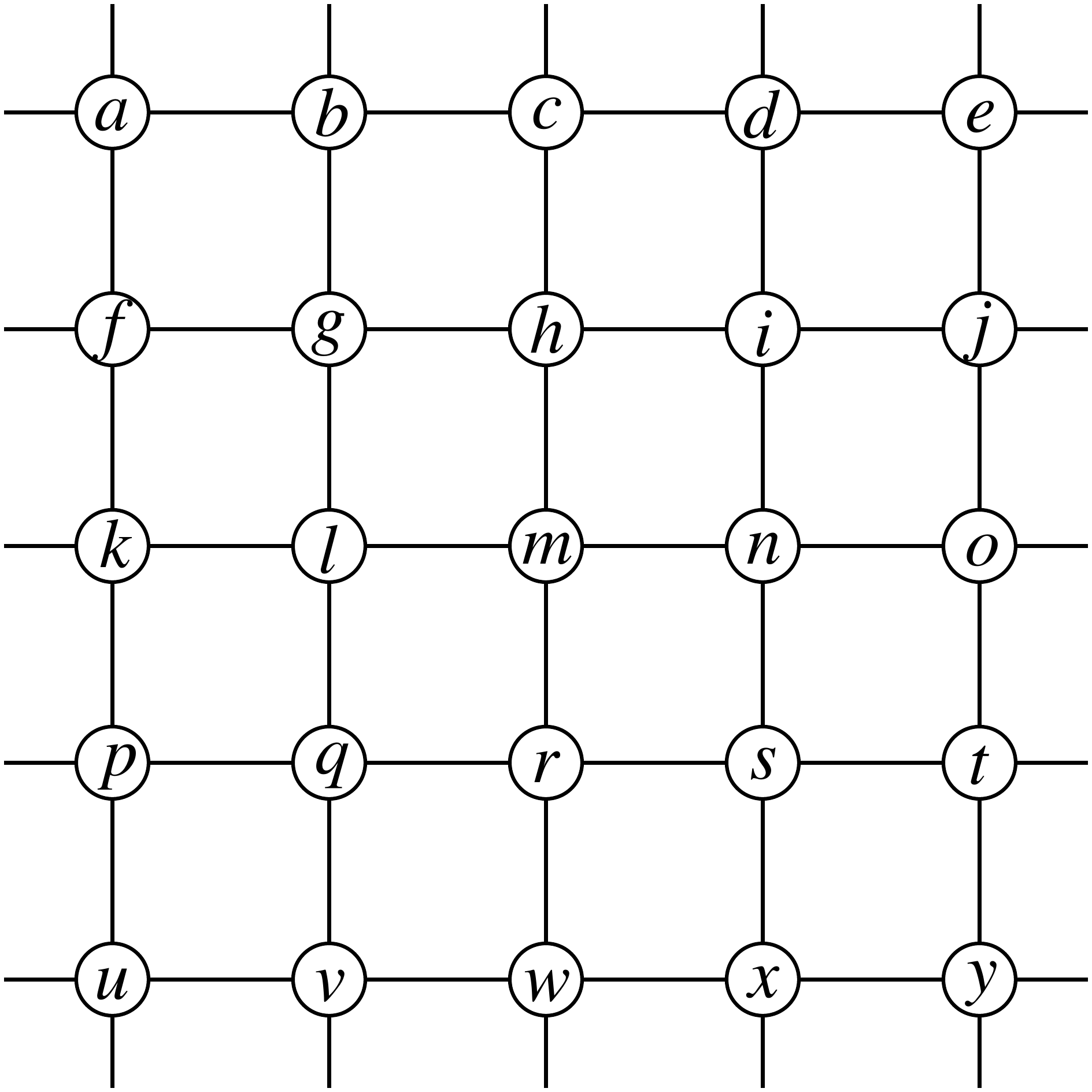}
  \end{center}
  \caption{
    \label{fig:2Dsquare}
    The square lattice $G$ with periodic boundary conditions. There are
    $L$ (here $L=5$) vertices on each boundary line, and the total number
    of vertices is $N=L \times L$.
  }
\end{figure}

Let us consider a periodic square lattice $G$ of width $L$ containing
$N = L \times L$ vertices, see Fig.~\ref{fig:2Dsquare} (the
numerical results shown in Fig.~\ref{fig:results} and 
Fig.~\ref{fig:magresults} correspond to $L=\infty$). 
Each vertex $m \in \{1, 2, \ldots, N\}$ of this lattice has a spin state
$\sigma_m \in \{-1, +1\}$ and it interacts with its four nearest neighboring
vertices. The interaction between two vertices $m$ and $n$ is represented by
an edge in the lattice and this edge is denoted as $\langle m, n\rangle$ in
our following discussions. The set formed by all the nearest neighboring
vertices of vertex $m$ is denoted as $\partial m$, i.e.,
$\partial m \equiv \{ n\, :\, \langle m, n\rangle \in G\}$.
For the particular example of Fig.~\ref{fig:2Dsquare}, 
$\partial m = \{l, h, n, r\}$ and $\partial n = \{m, i, o, s\}$. In addition,
we denote by $\partial m\backslash n$ the set obtained by deleting vertex $n$
from the set $\partial m$, e.g., $\partial m\backslash n=\{l,h,r\}$ and
$\partial n \backslash m = \{i, o, s\}$.

We denote a microscopic spin configuration of the whole lattice $G$ as
$\underline{\sigma}$, that is,
$\underline{\sigma} \equiv \{\sigma_1, \sigma_2, \ldots, \sigma_N\}$.
The energy for each of the $2^N$ possible microscopic configurations is
defined as
\begin{equation}
  \label{eq:energy}
  E(\underline{\sigma}) = - \sum\limits_{i \in G} h^0_i \sigma_i -
  \sum\limits_{\langle i, j\rangle \in G} J_{i j} \sigma_i \sigma_j \; ,
\end{equation}
where $h^0_i$ is the local external field on vertex $i$, and $J_{i j}$ is the
spin coupling constant of the edge $\langle i , j\rangle$.
In the limiting case of $J_{i j} = +J$ for all the edges, this model
is the ferromagnetic Ising model \cite{Ising-1925}. In the other limiting case
of the Edwards-Anderson spin glass model, each edge coupling constant
$J_{i j}$ is set to be $+J$ or $-J$ with equal probability and independently of
all the other coupling constants~\cite{Edwards-Anderson-1975}. In the numerical
calculations of this paper we choose the energy unit to be $J$, which is 
equivalent to setting $J=1$.

Let us denote by $\mathcal{S}$ a macroscopic equilibrium state of the
system at a given temperature $T$. When $T$ is sufficiently high the system
has only a single macroscopic state, then $\mathcal{S}$ contains
all the $2^N$ microscopic configurations. At certain critical temperature value
$T_c$ an ergodicity-breaking transition may occur in the configuration space
of the system, then the system at $T< T_c$ has two or even many macroscopic
states, each of which containing a subset of the $2^N$ microscopic
configurations~\cite{Huang-1987}
that are mutually reachable through a chain of local spin flips.

%%% figure 1

\section{The Bethe-Peierls approximation and the belief-propagation equation}
\label{sec:BP}

We now briefly review the BP method.
Within a macroscopic equilibrium state $\mathcal{S}$, the marginal probability
distribution $q_m(\sigma_m)$ for the spin state of a single vertex $m$ is 
defined as
\begin{equation}
  \label{eq:qm}
  q_m(\sigma) =
  \frac{\sum_{\underline{\sigma}}^\prime \delta_{\sigma_m}^{\sigma} e^{-\beta
      E(\underline{\sigma})}}
     {\sum_{\underline{\sigma}}^\prime e^{-\beta E(\underline{\sigma})}} \; ,
\end{equation}
where $\delta_\sigma^{\tilde{\sigma}}$ is the Kronecker symbol such that
$\delta_\sigma^{\tilde{\sigma}}=1$ if $\sigma=\tilde{\sigma}$ and 
$\delta_{\sigma}^{\tilde{\sigma}}=0$ if $\sigma \neq \tilde{\sigma}$;
$\beta \equiv 1/ T$ is the inverse temperature; the superscript ${^\prime}$
of the summation symbol means that the summation is over all the microscopic
configurations $\underline{\sigma}$ belonging to the macroscopic state
$\mathcal{S}$.

\begin{figure}
  \begin{center}
    \includegraphics[width=0.3\textwidth]{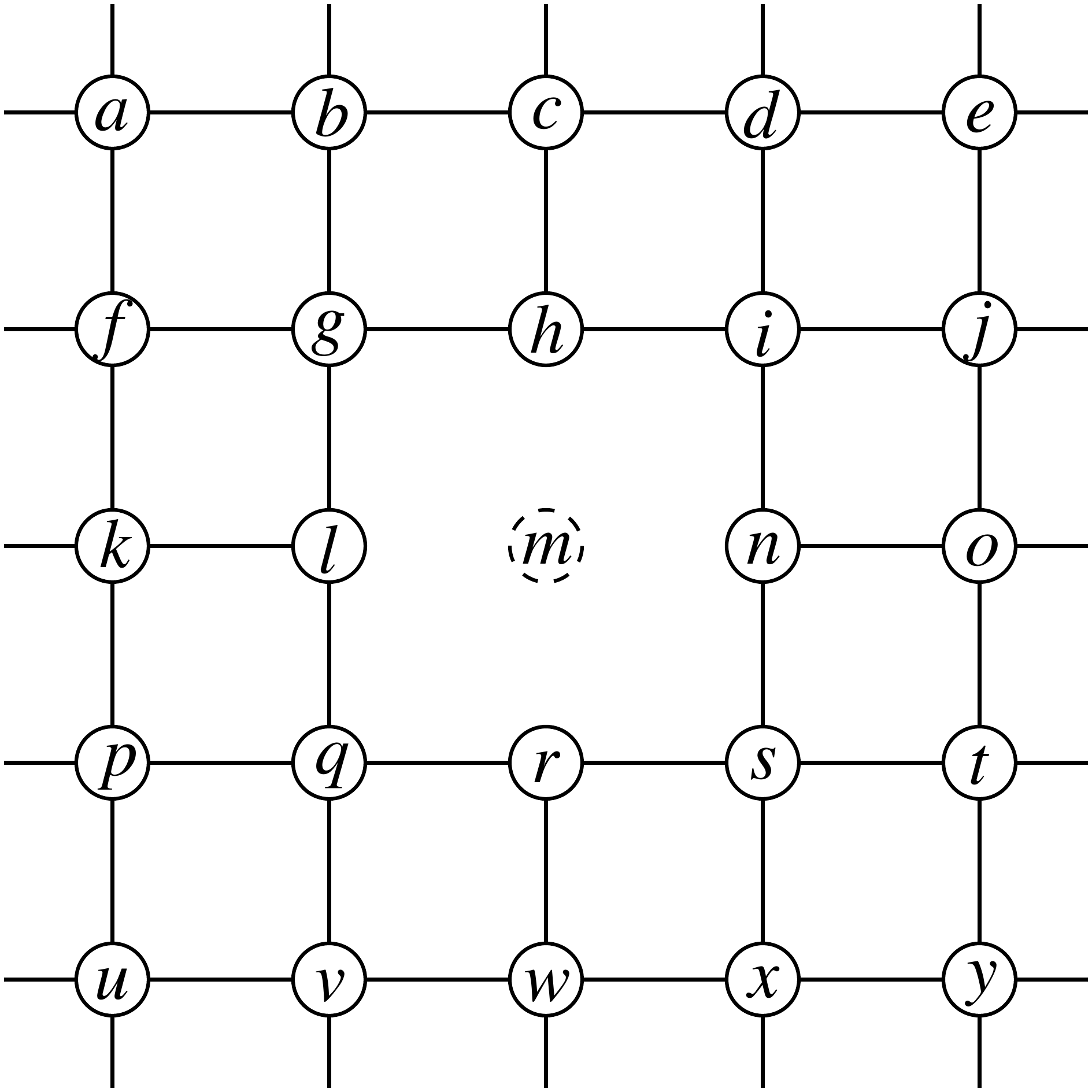}
  \end{center}
  \caption{
    \label{fig:2DsquareM}
    The square lattice $G_{\backslash m}$ obtained by deleting vertex $m$
    (and all its attached edges) from the lattice $G$ of
    Fig.~\ref{fig:2Dsquare}. Such a lattice is referred to as a cavity
    lattice.    
  }
\end{figure}

Since vertex $m$ interacts only with the vertices in $\partial m$,
we divide the total energy $E(\underline{\sigma})$ into two parts:
\begin{equation}
  \label{eq:energy2parts}
  E(\underline{\sigma}) = \Bigl[ -h^0_m \sigma_m - \sum\limits_{n\in \partial m}
    J_{m n} \sigma_m \sigma_n \Bigr] 
  + E_{\backslash m}(\underline{\sigma}_{\backslash m}) \; ,
\end{equation}
where $E_{\backslash m}(\underline{\sigma}_{\backslash m})$ is the total energy
of the cavity lattice $G_{\backslash m}$ formed by deleting vertex $m$ from the
original lattice $G$ (see Fig.~\ref{fig:2DsquareM}):
\begin{equation}
  E_{\backslash m}(\underline{\sigma}_{\backslash m}) = 
  - \sum\limits_{i\in G_{\backslash m}} h^0_i \sigma_i -
  \sum\limits_{\langle i, j\rangle \in G_{\backslash m}} J_{i j} \sigma_i \sigma_j \; ,
\end{equation}
and
$\underline{\sigma}_{\backslash m}\equiv\{\sigma_j\, :\, j\in G_{\backslash m}\}$.
After inserting Eq.~(\ref{eq:energy2parts}) into
Eq.~(\ref{eq:qm}), we obtain that
\begin{equation}
  \label{eq:qm2parts}
  q_m(\sigma) = \frac{
    e^{\beta h^0_m \sigma} \sum\limits_{\underline{\sigma}_{\partial m}}
    q_{\backslash m}(\underline{\sigma}_{\partial m})
    \prod\limits_{n\in \partial m} e^{\beta \sigma J_{m n} \sigma_n}
  }
  {
    \sum\limits_{\sigma_m} e^{\beta h^0_m \sigma_m} 
    \sum\limits_{\underline{\sigma}_{\partial m}}
    q_{\backslash m}(\underline{\sigma}_{\partial m})
    \prod\limits_{n\in \partial m} e^{\beta \sigma_m J_{m n} \sigma_n} 
  } \; ,
\end{equation}
where $\underline{\sigma}_{\partial m} \equiv \{\sigma_n : n \in \partial m\}$
denotes a microscopic spin configuration of the vertices in set $\partial m$,
and $q_{\backslash m}(\underline{\sigma}_{\partial m})$ is the
probability distribution of $\underline{\sigma}_{\partial m}$
within the macroscopic equilibrium state $\mathcal{S}$ of the cavity lattice
$G_{\backslash m}$:
\begin{equation}
  q_{\backslash m}(\underline{\sigma}_{\partial m}) \equiv
  \frac{\sum_{\underline{\tilde{\sigma}}_{\backslash m}}^{\prime}
    e^{ - \beta E_{\backslash m}(\underline{\tilde{\sigma}}_{\backslash m})}
    \prod_{n\in \partial m} \delta_{\tilde{\sigma}_n}^{\sigma_n}
  }
       {
         \sum_{\underline{\tilde{\sigma}}_{\backslash m}}^{\prime}
         e^{-\beta E_{\backslash m}(\underline{\tilde{\sigma}}_{\backslash m})}
       } \; .
\end{equation}
%

%%% figure 2

Since vertex $m$ is absent in the cavity lattice $G_{\backslash m}$, one expects
that the correlations among the vertices of set $\partial m$ are much weaker 
in $G_{\backslash m}$ than in the original lattice $G$. Following the idea of
Bethe and Peierls \cite{Bethe-1935,Peierls-1936a}, let us neglecting all the
remaining correlations among the vertices of $\partial m$ in $G_{\backslash m}$
and approximate $q_{\backslash m}(\underline{\sigma}_{\partial m})$ by the
following factorized form:
\begin{equation}
  \label{eq:BethePeierlsApprox}
  q_{\backslash m}(\underline{\sigma}_{\partial m}) \approx
  \prod\limits_{n\in \partial m} q_{n\backslash m}(\sigma_n) \; ,
\end{equation}
where $q_{n\backslash m}(\sigma_n)$ is the marginal probability distribution
of the spin state of vertex $n$ in the cavity lattice $G_{\backslash m}$.
Inserting Eq.~(\ref{eq:BethePeierlsApprox}) into Eq.~(\ref{eq:qm2parts}) we
obtain the following approximate expression for $q_m(\sigma_m)$:
\begin{equation}
  \label{eq:qmBP}
  q_m(\sigma) = \frac{
    e^{\beta h^0_m \sigma} \prod\limits_{n \in \partial m}
    \Bigl[ \sum\limits_{\sigma_n} e^{\beta  \sigma J_{m n} \sigma_n}
      q_{n\backslash m}(\sigma_n) \Bigr]
  }
  {
    \sum\limits_{\sigma_m}
    e^{\beta h^0_m \sigma_m} \prod\limits_{n \in \partial m}
    \Bigl[ \sum\limits_{\sigma_n} e^{\beta  \sigma_m J_{m n} \sigma_n}
      q_{n\backslash m}(\sigma_n) \Bigr]
  } \; .
\end{equation}

Similar to Eq.~(\ref{eq:qmBP}), we can apply the Bethe-Peierls approximation
on the cavity lattice $G_{\backslash m}$ to compute the marginal probability
distribution $q_{n\backslash m}(\sigma_m)$ of vertex $n$:
\begin{equation}
  \label{eq:BP}
  q_{n\backslash m}(\sigma)
  = \frac{
    e^{\beta h^0_n \sigma} \prod\limits_{i \in \partial n\backslash m}
    \Bigl[ \sum\limits_{\sigma_i} e^{\beta  \sigma J_{n i} \sigma_i}
      q_{i\backslash n}(\sigma_i) \Bigr]
  }
  {
    \sum\limits_{\sigma_n}
    e^{\beta h^0_n \sigma_n} \prod\limits_{i \in \partial n\backslash m}
    \Bigl[ \sum\limits_{\sigma_i} e^{\beta  \sigma_n J_{n i} \sigma_i}
      q_{i\backslash n}(\sigma_i) \Bigr]
  } \; .
\end{equation}
The above equation is referred to as a belief-propagation equation in
the literature \cite{Pearl-1988}. The BP equation is a self-consistent
equation. We can iterate Eq.~(\ref{eq:BP}) on all the edges of the lattice
$G$ and, if this iteration reaches a fixed point, then use Eq.~(\ref{eq:qmBP})
to compute the mean spin value of any given vertex $m$ in the lattice.

The above-mentioned mean field theory is very successful in quantitatively
predicting the properties of spin models on random finite-connectivity graphs
\cite{ZhouBook}.
However, when applied on the square-lattice Ising model with no external field,
it predicts a transition between the paramagnetic phase and the ferromagnetic
phase at the critical inverse temperature $\beta \approx 0.3466$, which is
considerably lower than the exact value
$\beta_c = \ln (1+\sqrt{2})/2 \approx 0.4407$ 
\cite{Kramers-Wannier-1941a,Kramers-Wannier-1941b}, see Fig.~\ref{fig:results}. 
For the Edwards-Anderson spin glass model on the periodic square lattice
(again with no external field), the paramagnetic solution of the BP equation
(\ref{eq:BP}) becomes unstable 
  as $\beta$ exceeds certain threshold value
  $\beta_c(L)$ which is a decreasing function of
  lattice size $L$ and $\lim_{L\rightarrow \infty} \beta(L) \approx 0.370$
  \cite{Zhou-Wang-2012};  BP converges to a non-paramagnetic fixed point 
  at $\beta$ slightly beyond $\beta_c(L)$, but it fails to
  converge at $\beta > 0.66$ (see, for example, 
  \cite{LageCastellanos-etal-2011}).
  These latter results are contradicting
with the strong numerical evidence
\cite{Morgenstern-Binder-1980,Saul-Kardar-1993,Houdayer-2001,Joerg-etal-2006,Thomas-Middleton-2009,Toldin-Pelissetto-Vicari-2010,Thomas-Huse-Middleton-2011,Toldin-Pelissetto-Vicari-2011}
that the two-dimensional Edwards-Anderson model is in the paramagnetic phase 
at any finite $\beta$.

The mean-field equations (\ref{eq:qmBP}) and (\ref{eq:BP}) are not accurate in
treating lattice spin models. We now develop a loop-corrected belief
propagation numerical scheme to better considering the complicated effect of
short loops.

\section{Loop-corrected belief-propagation equation}

We notice that, due to the abundance of short loops, the naive BP equations
(\ref{eq:qmBP}) and (\ref{eq:BP}) generate a spurious self-field on each vertex
of the lattice. By definition the probability distribution
$q_{n\backslash m}(\sigma_n)$ in Eq.~(\ref{eq:qmBP}) is completely independent
of vertex $m$, but if we use Eq.~(\ref{eq:BP}) then $q_{n\backslash m}(\sigma_n)$
will be strongly affected by $m$. To explain this point by an example, let us
consider the path $m$--$h$--$i$--$n$ in Fig.~\ref{fig:2DsquareM}:
$q_{n\backslash m}(\sigma_n)$ depends on $q_{i\backslash n}(\sigma_i)$, which in
turn depends on $q_{h\backslash i}(\sigma_h)$, which in turn depends on
$q_{m\backslash h}(\sigma_m)$. Similarly, other short paths between vertex $n$
and vertex $m$ will bring additional dependence of
$q_{n\backslash m}(\sigma_n)$ on the `deleted' vertex $m$. Since all the input
probability distributions to vertex $m$ in Eq.~(\ref{eq:qmBP}) actually are
affected by vertex $m$, the resulting marginal probability distribution
$q_m(\sigma_m)$ contains the self-field of vertex $m$ to itself. This
self-field effect is not real but is an artifact of the naive BP equation
(\ref{eq:BP}).

%%% figure 3
\begin{figure}
  \begin{center}
    \includegraphics[angle=270,width=0.49\textwidth]{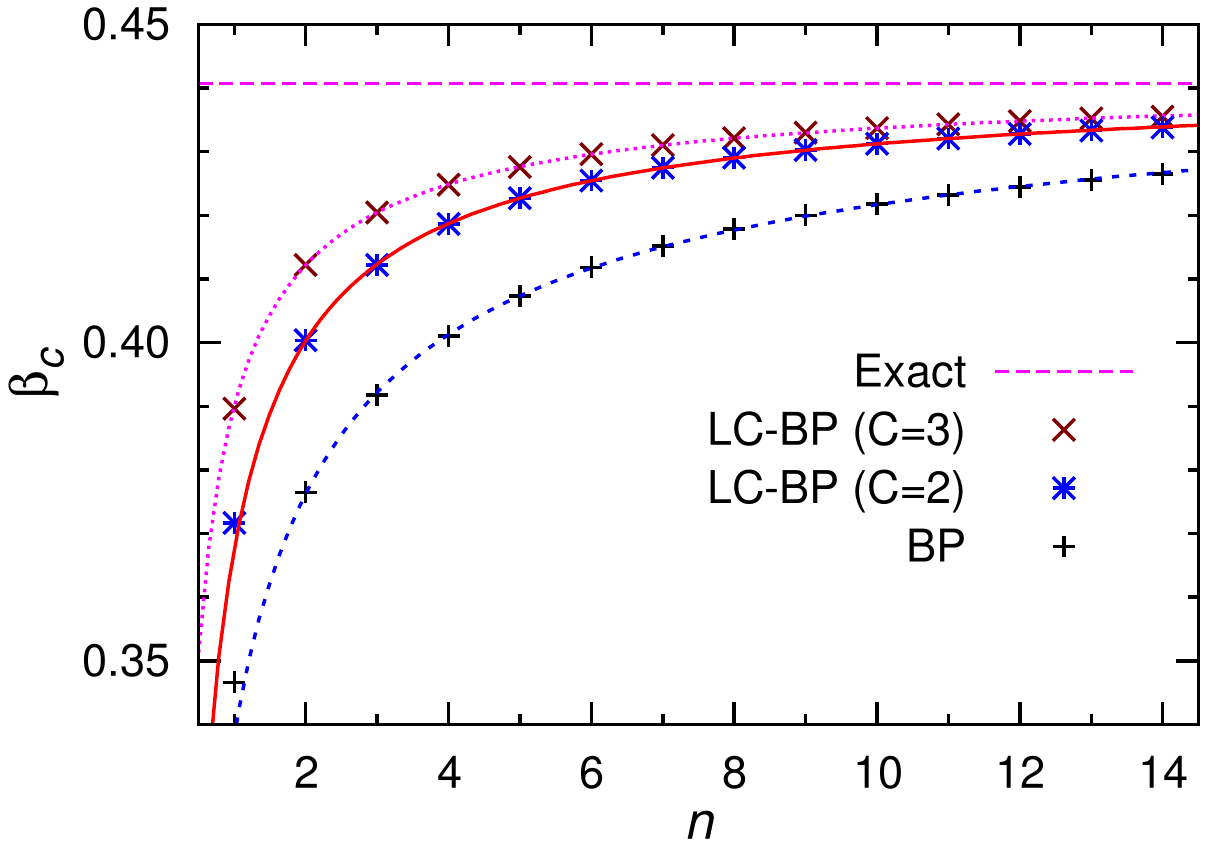}
  \end{center}
  \caption{
    \label{fig:results}
    The inverse temperature $\beta_c$ at the ferromagnetic phase transition
    point of the square-lattice Ising model (no external field). The results
    obtained by the belief-propagation equation (BP, plus symbols) and those
    obtained by the loop-corrected belief-propagation equation (LC-BP) with
    memory capacity $C=2$ (star symbols) and memory capacity $C=3$ (cross 
    symbols) are compared with the exact value $\beta_c \approx 0.4407$ (marked
    by the horizontal dashed line). Each square region of BP and LC-BP
    contains $n\times n$ vertices, with $n$ being the number of
    vertices along one boundary line of the square region.
    We can fit the data by the function 
    $\beta_c = \beta_c^\infty - c\, n^{-\gamma}$, with
    $\beta_c^\infty = 0.4490$, $c=0.1109$ and $\gamma=0.6075$ (for BP,
    bottom dashed curve), $\beta_c^\infty = 0.4421$, $c=0.0746$ and
    $\gamma=0.8357$ (for LC-BP at $C=2$, middle solid curve),
    and $\beta_c^\infty = 0.4417$, $c=0.0517$ and $\gamma=0.8071$ (for
    LC-BP at $C=3$, top dotted curve).
  }
\end{figure}

\begin{figure}
  \begin{center}
    \includegraphics[width=0.3\textwidth]{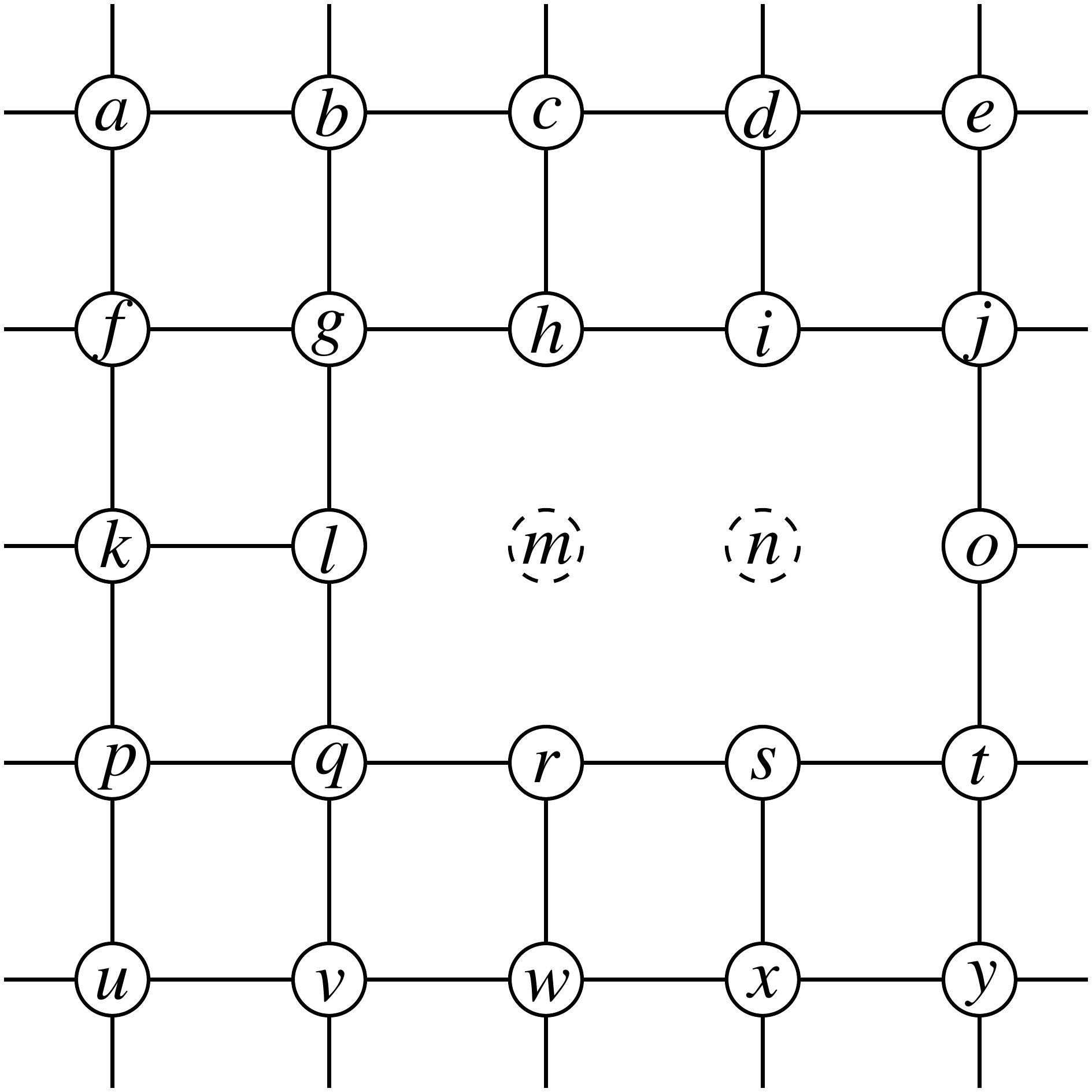}
  \end{center}
  \caption{
    \label{fig:2DsquareMN}
    The cavity square lattice $G_{\backslash \{m,n\}}$ obtained by deleting
    vertices $m$ and $n$ (and all the attached edges) from the lattice $G$ of 
    Fig.~\ref{fig:2Dsquare}.
  }
\end{figure}

We need to modify Eq.~(\ref{eq:BP}) to remove this spurious self-field effect.
Actually, if we strictly follow the Bethe-Peierls approximation, the expression
for the probability distribution $q_{n\backslash m}(\sigma_n)$ is not
Eq.~(\ref{eq:BP}) but the following:
\begin{equation}
  \label{eq:lcBP1}
  q_{n\backslash m}(\sigma_n)
  = \frac{
    e^{\beta h^0_n \sigma_n} \prod\limits_{i \in \partial n\backslash m}
    \Bigl[ \sum\limits_{\sigma_i} e^{\beta  \sigma_n J_{n i} \sigma_i}
      q_{i\backslash \{m,n\}}(\sigma_i) \Bigr]
  }
  {
    \sum\limits_{\sigma_n^\prime}
    e^{\beta h^0_n \sigma_n^\prime} \prod\limits_{i \in \partial n\backslash m}
    \Bigl[ \sum\limits_{\sigma_i} e^{\beta  \sigma_n^\prime J_{n i} \sigma_i}
      q_{i\backslash \{m,n\}}(\sigma_i) \Bigr]
  } \; ,
\end{equation}
where $q_{i\backslash \{m, n\}}(\sigma_i)$ is the marginal probability distribution
of vertex $i$'s spin state in the cavity lattice $G_{\backslash \{m,n\}}$ with
both vertex $m$ and $n$ being deleted (see Fig.~\ref{fig:2DsquareMN}).

In general, for any given vertex set $\phi$ and a vertex $n$ that is adjacent
to at least one vertex in this set $\phi$, we denote by
$q_{n\backslash \phi}(\sigma_n)$ the marginal probability distribution of
vertex $n$'s spin state in the cavity lattice $G_{\backslash \phi}$ obtained by 
deleting all the vertices of $\phi$ from the original lattice $G$. Under the
Bethe-Peierls approximation, this probability distribution can be determined
through
\begin{equation}
  \label{eq:lcBP2}
  q_{n\backslash \phi}(\sigma)
  = \frac{
    e^{\beta h^0_n \sigma} \prod\limits_{i \in \partial n\backslash \phi}
    \Bigl[ \sum\limits_{\sigma_i} e^{\beta  \sigma J_{n i} \sigma_i}
      q_{i\backslash \{\phi,n\}}(\sigma_i) \Bigr]
  }
  {
    \sum\limits_{\sigma_n}
    e^{\beta h^0_n \sigma_n} \prod\limits_{i \in \partial n\backslash \phi}
    \Bigl[ \sum\limits_{\sigma_i} e^{\beta  \sigma_n J_{n i} \sigma_i}
      q_{i\backslash \{\phi, n\}}(\sigma_i) \Bigr]
  } \; ,
\end{equation}
where $\partial n \backslash \phi \equiv \partial n - \phi \cap \partial n$ 
denotes the vertex set obtained by deleting all the vertices of $\partial n$
that are also belonging to set $\phi$, 
and $\{\phi, n\} \equiv \phi \cup \{n\}$ is the vertex set obtained by adding
vertex $n$ to set $\phi$.

%%% figure 4

Equations (\ref{eq:qmBP}), (\ref{eq:lcBP1}) and (\ref{eq:lcBP2}) form a
hierarchical series of self-consistent equations and we refer them 
collectively as the loop-corrected belief-propagation equation.
For practical applications we have to make a cutoff to this message-passing
hierarchy, so that a closed set of equations can be obtained and can be
iterated numerically.

In the remaining part of this paper we mainly consider the simplest nontrivial
cutoff by requiring that the vertex set $\phi$ of the cavity probability 
distribution $q_{n\backslash \phi}$ of any vertex $n$ can contain at most
two vertices (i.e., memory capacity $C=2$). Under this additional restriction,
then for the two vertices $l$ and $r$ in Fig.~\ref{fig:2DsquareMN} we have
\begin{subequations}
  \label{eq:lcBPmax2}
  \begin{align}
    q_{l\backslash \{m,n\}}(\sigma_l) & \propto
    e^{\beta h^0_l \sigma_l} \bigl[\sum{_{\sigma_g}}
      e^{\beta \sigma_l J_{l g}\sigma_g} q_{g\backslash \{l,m\}}(\sigma_g)\bigr]
    \nonumber \\
    & \quad \quad \times \bigl[\sum{_{\sigma_k}}
      e^{\beta \sigma_l J_{l k}\sigma_k} q_{k\backslash \{l,m\}}(\sigma_k)\bigr]
    \nonumber \\
    & \quad \quad  \times \bigl[\sum{_{\sigma_q}}
      e^{\beta \sigma_l J_{l q}\sigma_q} q_{q\backslash \{l,m\}}(\sigma_q)\bigr]
    \; , \label{eq:lcBPmax2a}
    \\
    q_{r\backslash \{m,n\}}(\sigma_r) & \propto
    e^{\beta h^0_r \sigma_r} \bigl[\sum{_{\sigma_q}}
      e^{\beta \sigma_r J_{r q}\sigma_q} q_{q\backslash \{r, m\}}(\sigma_q)\bigr]
    \nonumber \\
    & \quad \quad \times\bigl[\sum{_{\sigma_w}}
      e^{\beta \sigma_r J_{r w}\sigma_w} q_{w\backslash \{r, m\}}(\sigma_w)\bigr]
    \nonumber \\
    & \quad \quad \times \bigl[\sum{_{\sigma_s}}
      e^{\beta \sigma_r J_{r s}\sigma_s} q_{s\backslash \{r, n\}}(\sigma_s)\bigr]
    \; . \label{eq:lcBPmax2b}
    \end{align}
\end{subequations}
We consider $q_{s\backslash \{r, n\}}(\sigma_s)$ instead of
$q_{s\backslash \{r,m\}}(\sigma_s)$
in the last line of Eq.~(\ref{eq:lcBPmax2b})
because vertex $n$ has stronger influence
to vertex $s$ than vertex $m$.
The probability distribution $q_{s\backslash \{r, n\}}(\sigma_s)$
of Eq.~(\ref{eq:lcBPmax2b}) can be computed through
\begin{eqnarray}
  q_{s\backslash \{n,r\}}(\sigma_s)  &\propto &
  e^{\beta h^0_s \sigma_s} \bigl[\sum{_{\sigma_t}}
    e^{\beta \sigma_s J_{s t}\sigma_t} q_{t\backslash \{s, n\}}(\sigma_t)\bigr]
  \nonumber \\
  & &  \times
  \bigl[\sum{_{\sigma_x}}
    e^{\beta \sigma_s J_{s x}\sigma_x} q_{x\backslash \{r, s\}}(\sigma_x)\bigr]
  \; .
  \label{eq:lcBPmax2c}
\end{eqnarray}

When we apply Eqs.~(\ref{eq:lcBPmax2}) and (\ref{eq:lcBPmax2c}) to the
square-lattice Ising model, we obtain a critical inverse temperature
$\beta_c\approx 0.3716$ for the ferromagnetic phase transition, which is
considerably better than the prediction of the naive BP, 
see Fig.~\ref{fig:results}. This is an encouraging result. We can further
improve the performance of the loop-corrected BP mean field theory by 
allowing the set
$\phi$ of deleted vertices in Eq.~(\ref{eq:lcBP2}) to contain three or even
more vertices. For example if the memory capacity is set to $C=3$ the
value of $\beta_c$ estimated for the ferromagnetic Ising model increases to
$\beta_c \approx 0.3896$ (see Fig.~\ref{fig:results}).

The mean magnetization $\langle \sigma_m \rangle$ of vertex $m$ and
the mean spin correlation $\langle \sigma_m \sigma_n \rangle$ between
vertex $m$ and $n$ are estimated through the following equations:
\begin{subequations}
  \begin{align}
    \langle \sigma_m \rangle & =
    \sum\limits_{\sigma_m} \sigma_m q_m(\sigma_m) \; ,
    \label{eq:sigmaM} 
    \\
    \langle \sigma_m \sigma_n \rangle & =
    \frac{\sum\limits_{\sigma_m , \sigma_n}
      \sigma_m \sigma_n e^{\beta J_{m n} \sigma_m \sigma_n}
      q_{m\backslash n}(\sigma_m)
      q_{n\backslash m}(\sigma_n)}
         {
           \sum\limits_{\sigma_m, \sigma_n}
           e^{\beta J_{m n} \sigma_m \sigma_n}
           q_{m\backslash n}(\sigma_m)
           q_{n\backslash m}(\sigma_n)}
         \; . 
         \label{eq:correlate}
  \end{align}
\end{subequations}
The mean energy of the whole system is then 
\begin{equation}
  \label{eq:meanE}
  \langle E \rangle = - \sum\limits_{m=1}^{N} h_m^0 \langle \sigma_m \rangle
  - \sum\limits_{(m,n)\in G} J_{m n} \langle \sigma_m \sigma_n \rangle
  \; .
\end{equation}
$\langle E \rangle$ of course depends on the inverse 
temperature $\beta$, let us emphasize this dependence by
$\langle E \rangle_\beta$. 
The free energy $F(\beta)$ of the
system is related to the mean energy through 
$\langle E \rangle_\beta = \frac{{\rm d} (\beta F)}{{\rm d} \beta}$,
namely
\begin{equation}
  F(\beta) = \frac{1}{\beta} \int\limits_{0}^{\beta}
  \langle E \rangle_{\beta^\prime} {\rm d} \beta^\prime
  -\frac{1}{\beta} N \ln 2 \; .
\end{equation}

\section{Loop-corrected belief propagation at the region graph level}

In essence, the loop-corrected BP mean field theory of the preceding section
tries to completely eliminate the effect of a deleted vertex $m$ to the cavity
lattice $G_{\backslash m}$ through the BP hierarchy Eqs.~(\ref{eq:lcBP1}) and
(\ref{eq:lcBP2}). But the loop-corrected BP hierarchy is also based on the
Bethe-Peierls approximation and it does not consider any of the short-range
correlations that are discarded from this approximation (e.g., the correlations
among the vertices $l$, $h$, $n$, and $m$ in the cavity graph $G_{\backslash m}$ 
of Fig.~\ref{fig:2DsquareM}).
To take into account more short-range correlations, we follow the work of
Zhou and Wang \cite{Zhou-Wang-2012} and construct the loop-corrected BP
equation at the coarse-grained region graph level.

\begin{figure}
  \begin{center}
    \includegraphics[width=0.3\textwidth]{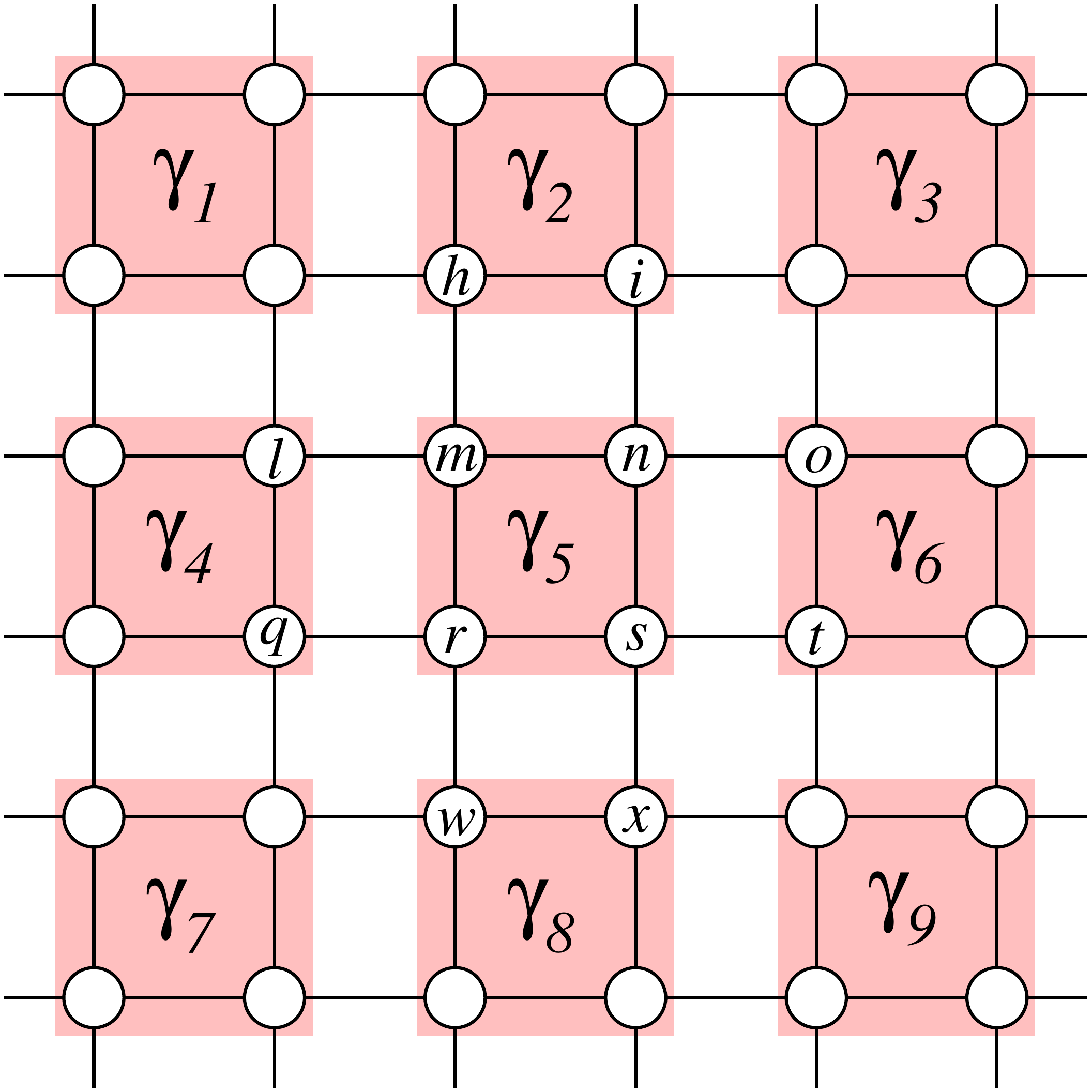}
  \end{center}
  \caption{
    \label{fig:2Dregion}
    The square lattice coarse-grained as a region graph $\mathcal{R}$.
    Each square region ($\gamma_1, \gamma_2, \ldots$) contains 
    $n \times n$ vertices and all the interactions within these vertices
    ($n=2$ in this particular example). Two nearest-neighboring regions
    interact through $n$ edges.
  }
\end{figure}

In the example of the square lattice, we completely cover the vertices of the 
whole lattice by a set of square regions without any overlap between the
regions. Each square region contains $n \times n$ vertices and all the
interaction edges within these vertices, see Fig.~\ref{fig:2Dregion}. 
Two neighboring regions interact with each other through the $n$ edges in
between, and they are therefore considered as being connected at the region
level. The region graph $\mathcal{R}$ constructed in this way, with each vertex
representing a local square domain of $n\times n$ vertices,
has the same topology as the original square lattice $G$. 

The loop-corrected BP hierarchy can then be obtained for this region graph
 $\mathcal{R}$.
Consider the region $\gamma_5$ of Fig.~\ref{fig:2Dregion} as an example. Let
us define $q_{\gamma_5 \backslash \gamma_2}(\sigma_m, \sigma_n)$ as the probability
of vertex $m$ taking spin value $\sigma_m$ and vertex $n$ taking spin value
$\sigma_n$ in the cavity region graph $\mathcal{R}_{\backslash \gamma_2}$ obtained
by deleting region $\gamma_2$ from $\mathcal{R}$. Other joint probability
distributions can be defined in a similar way, e.g., 
$q_{\gamma_5 \backslash \{\gamma_1, \gamma_2\}}(\sigma_m, \sigma_n)$ is the joint
probability distribution of $\sigma_m$ and $\sigma_n$ in the cavity region
graph $\mathcal{R}_{\backslash \{\gamma_1, \gamma_2\}}$ (with regions $\gamma_1$
and $\gamma_2$ being deleted).
If we restrict the set $\phi$ of deleted regions in memory to containing two
regions at most (i.e., memory capacity $C=2$), we obtain that
\begin{subequations}
  \label{eq:lcBPregion}
  \begin{align}
   & q_{\gamma_5 \backslash \gamma_2}(\sigma_m, \sigma_n)  \propto 
    \sum{_{\sigma_r, \sigma_s}} e^{-\beta E_{\gamma_5}} \nonumber \\
    & \quad \quad \times \bigl[
      \sum{_{\sigma_l, \sigma_q}} e^{-\beta E_{\gamma_4 \gamma_5}}
      q_{\gamma_4\backslash \{\gamma_2, \gamma_5\}}(\sigma_l, \sigma_q) \bigr]
    \nonumber \\
   &  \quad \quad \times \bigl[
      \sum{_{\sigma_w, \sigma_x}} e^{-\beta E_{\gamma_8 \gamma_5}}
      q_{\gamma_8\backslash \{\gamma_2, \gamma_5\}}(\sigma_w, \sigma_x) \bigr]
    \nonumber \\
    & \quad\quad \times \bigl[
      \sum{_{\sigma_t, \sigma_o}} e^{-\beta E_{\gamma_6 \gamma_5}}
      q_{\gamma_6\backslash \{\gamma_2, \gamma_5\}}(\sigma_t, \sigma_o) \bigr]
    \; ,
    \\
    & q_{\gamma_5 \backslash\{\gamma_1, \gamma_2\}}(\sigma_m, \sigma_n)  \propto 
    \sum{_{\sigma_r, \sigma_s}} e^{-\beta E_{\gamma_5}} \nonumber \\
    & \quad \quad \times \bigl[
      \sum{_{\sigma_l, \sigma_q}} e^{-\beta E_{\gamma_4 \gamma_5}}
      q_{\gamma_4\backslash \{\gamma_1, \gamma_5\}}(\sigma_l, \sigma_q) \bigr]
    \nonumber \\
    &  \quad \quad \times \bigl[
      \sum{_{\sigma_w, \sigma_x}} e^{-\beta E_{\gamma_8 \gamma_5}}
      q_{\gamma_8\backslash \{\gamma_2, \gamma_5\}}(\sigma_w, \sigma_x) \bigr]
    \nonumber \\
    & \quad\quad \times \bigl[
      \sum{_{\sigma_t, \sigma_o}} e^{-\beta E_{\gamma_6 \gamma_5}}
      q_{\gamma_6\backslash \{\gamma_2, \gamma_5\}}(\sigma_t, \sigma_o) \bigr]
    \; ,
    \\
    & q_{\gamma_5 \backslash \{\gamma_2,\gamma_4\}}(\sigma_m, \sigma_n)  \propto 
    \sum{_{\sigma_r, \sigma_s}} e^{-\beta E_{\gamma_5}} \nonumber \\
    &  \quad \quad \times \bigl[
      \sum{_{\sigma_w, \sigma_x}} e^{-\beta E_{\gamma_8 \gamma_5}}
      q_{\gamma_8\backslash \{\gamma_4, \gamma_5\}}(\sigma_w, \sigma_x) \bigr]
    \nonumber \\
    & \quad\quad \times \bigl[
      \sum{_{\sigma_t, \sigma_o}} e^{-\beta E_{\gamma_6 \gamma_5}}
      q_{\gamma_6\backslash \{\gamma_2, \gamma_5\}}(\sigma_t, \sigma_o) \bigr]
    \; .
  \end{align}
\end{subequations}
In the above expressions, the quantity $E_{\gamma}$ denotes the internal energy 
of a region $\gamma$, for example
\begin{eqnarray}
  & & E_{\gamma_5}(\sigma_m, \sigma_n, \sigma_r, \sigma_s) =
  -h^0_m \sigma_m - h^0_n \sigma_n - h^0_r \sigma_r - h^0_s \sigma_s \nonumber \\
  & & \quad 
  -J_{m n} \sigma_m \sigma_n - J_{n s} \sigma_n \sigma_s 
  - J_{r s} \sigma_r \sigma_s
-J_{m r} \sigma_m \sigma_r \; ,
\end{eqnarray}
and $E_{\gamma \gamma^\prime}$ is the interaction energy between region $\gamma$
and region $\gamma^\prime$, for example
\begin{equation}
  E_{\gamma_4 \gamma_5}(\sigma_l, \sigma_m, \sigma_q, \sigma_r) =
  -J_{l m} \sigma_l \sigma_m - J_{q r} \sigma_q \sigma_r \; .
\end{equation}
%

%%% figure 5

As Eq.~(\ref{eq:lcBPregion}) demonstrates, all the correlations within each
region are precisely considered by summing over all the $2^{n^2}$ microscopic
configurations of this region. In the practical implementation, the internal
state summation is achieved through a numerical scheme that is efficient both
in terms of computing time and in terms of needed memory (see Appendix A
for details). 
By increasing the region size $n$ we can include more and more short-range
correlations and achieve more precise quantitative predictions.

\begin{figure}
  \begin{center}
    \includegraphics[angle=270,width=0.49\textwidth]{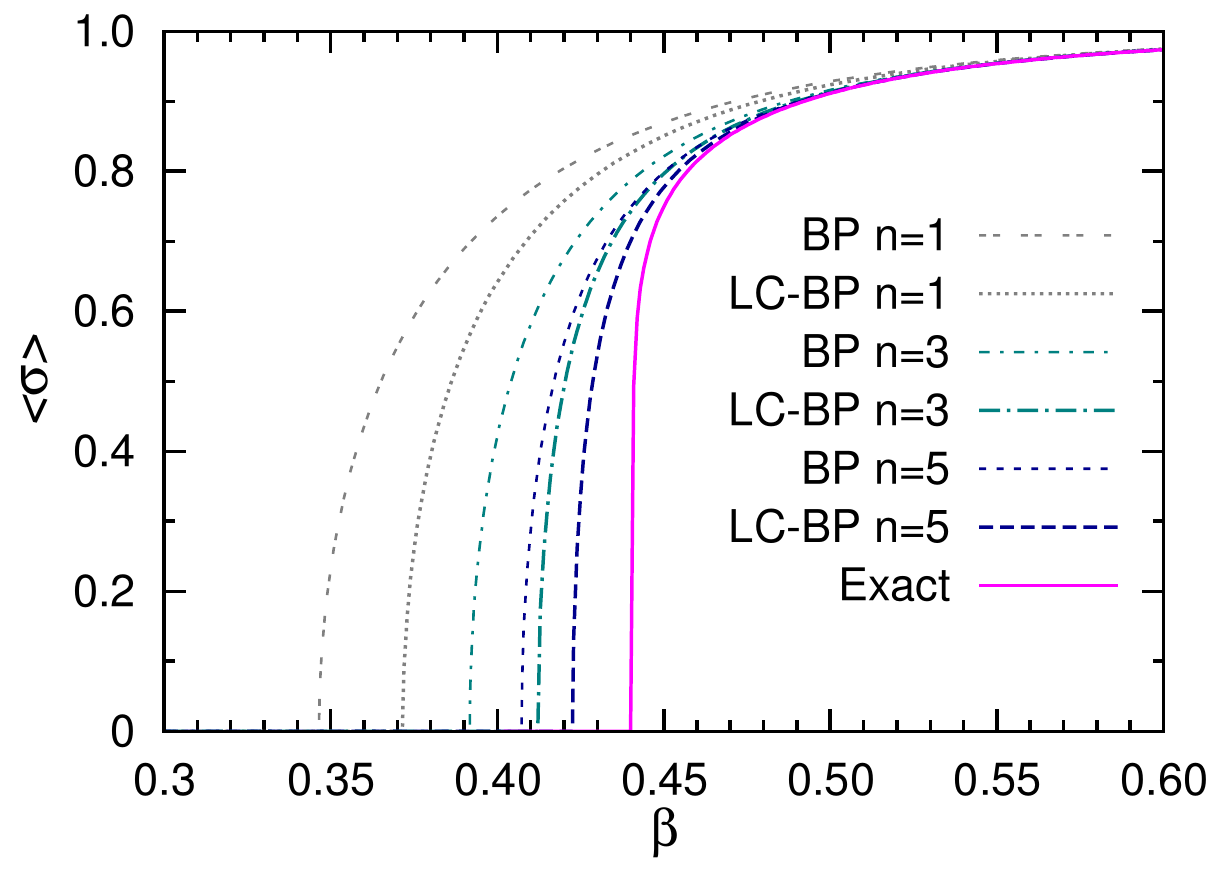}
  \end{center}
  \caption{
    \label{fig:magresults}
    The spontaneous magnetization (the mean spin value $\langle \sigma \rangle$
    of a vertex) of the square-lattice Ising model.
    The results obtained by belief propagation (BP) and those
    obtained by loop-corrected belief propagation (LC-BP) with memory
    capacity $C=2$ are shown together with the exact results (the solid line).
    Each square region of  BP and LC-BP
    contains $n\times n$ vertices, with $n =1$, $3$, or $5$.
  }
\end{figure}

For the two-dimensional Ising model we have compared in 
Fig.~\ref{fig:results} the results obtained by the conventional region-graph
BP of \cite{Zhou-Wang-2012} and those obtained by the present region-graph
loop-corrected BP. When the memory capacity is set to $C=2$ (the smallest
nontrivial value), the iteration process of loop-corrected BP demands the same
order of computational cost as that of BP, yet at each value of the 
square-region size $n$ the improvement of loop-corrected BP over BP is always
significant, suggesting that loop-corrected BP is a much better choice than
the naive BP for treating finite-dimensional lattice systems.
Figure~\ref{fig:magresults} compares the exact spontaneous magnetization of 
the square-lattice Ising model with the predictions obtained by BP and 
LC-BP ($C=2$).  At each value of the region sizes used ($n=1$, $n=3$,
or $n=5$) the improvement of LC-BP over BP is again significant.

It also appears that loop-corrected BP (with memory capacity $C=2$)
outperforms the GBP method of Yedidia and coworkers 
\cite{Yedidia-Freeman-Weiss-2005}. When the square-region size is set
to $n=2$, GBP predicts the critical inverse temperature of ferromagnetic
phase transition to be $\beta_c \approx 0.4126$ \cite{Wang-Zhou-2013}; a
slightly better result is achieved by the loop-corrected BP method at
square-region size $n^\prime =2 n = 4$, which reports a 
value of $\beta_c \approx 0.4186$. The GBP with square-region size $n=4$
predicts a value of $\beta_c \approx 0.429$ \cite{Wang-Zhou-2013}; this result
is marched by the loop-corrected BP at square-region size
$n^\prime = 2 n =8$, which reports a value of $\beta_c \approx 0.4290$.
We might therefore conjecture that GBP at square-region size $n$ and 
loop-corrected BP at square-region size $n^\prime = 2 n$ have comparable 
prediction power.
Under such an assumption we can then argue that loop-corrected BP will be a
better choice than GBP: (1) the iteration process of GBP is much more
complicated than that of loop-corrected BP; and (2) the required computer
storage space of a GBP message is of order $O(2^{n^2/2})$, making it unpractical
to set the square-region size $n\geq 6$; (3) the required storage space of
a loop-corrected BP message is only of order $O(2^n)$, so we can set the
square-region size to $n=20$ or even larger values.
  It should be pointed out that good performance of GBP can be
  achieved by increasing the size of the largest region one-dimensionally
  rather than two-dimensionally (see \cite{Kikuchi-Brush-1967}
  and \cite{Pelizzola-2005}). It will be helpful to perform
  a comparative study by implementing LC-BP also in such a non-symmetric way.
  We leave this point for future investigations.

We can further improve the performance of the loop-corrected BP method
by increasing the memory capacity $C$ (but at the cost of introducing many 
more cavity messages, see Appendix B). For the square-lattice Ising model,
the results obtained by loop-corrected BP at $C=3$ are also shown in 
Fig.~\ref{fig:results} to compare with the results obtained at $C=2$. 
We find that increasing $C$ from $C=2$ to $C=3$ does not bring a dramatic
improvement to the prediction of $\beta_c$. Considering the high computation
cost required for $C\geq 3$ (see Appendix B), if higher numerical precision
is needed, it is more practical to increase the square-region
size $n$ but to keep the memory capacity at $C=2$.

%%% figure 6

\section{Conclusion}

To summarize briefly, in this paper we described the main ideas of the
loop-corrected belief propagation method and carried out an initial performance
test on the square-lattice Ising model. The results in Fig.~\ref{fig:results}
and Fig.~\ref{fig:magresults}
clearly demonstrate that loop-corrected BP with memory capacity
$C=2$ is much superior to the naive BP method, which is equivalent to
loop-corrected BP with memory capacity $C=1$. The performance of loop-corrected
BP further improves as the memory capacity is increased to $C=3$ or even
larger values.

Our numerical results on the square-lattice Ising model also indicate that,
compared to the generalized belief propagation method of Yedidia 
{\it et al.} \cite{Yedidia-Freeman-Weiss-2005}, the loop-corrected
BP method (simply with memory capacity $C=2$) can achieve the same or even 
higher level of precision at much reduced computation cost.
In addition, we wish to point out another very important advantage of the
loop-corrected BP method: just as the survey propagation method is a natural
extension of the naive BP method \cite{Mezard-Montanari-2009,ZhouBook}, 
following the discussion of \cite{Zhou-Wang-2012}
we might extend loop-corrected BP into the loop-corrected survey 
propagation method to study disordered lattice models in the low-temperature
spin glass phase, where ergodicity of the configuration space is broken.

For the loop-corrected BP method really to be a helpful tool, it should be
capable of giving good quantitative predictions on single instances of 
disordered lattice models.  
The performance of loop-corrected BP on the square-lattice and cubic-lattice
spin glass models will be investigated and be reported in a forthcoming paper.

\section*{Acknowledgement}

Part of this work was carried out while one of the authors (HJZ) was visiting
the Physics Department of Zhejiang University. HJZ thanks Prof. Bo Zheng for
hospitality. This work was supported by the National Basic Research
Program of China (grant number 2013CB932804) and by the National Natural 
Science Foundation of China (grant numbers 11121403, 11175224, and 11225526).

Author contribution statement: HJZ, WMZ conceived research; HJZ performed
research and wrote the paper.

\begin{appendix}

  \section*{Appendix A: Message updating for a square region}

\begin{figure}
    \begin{center}
      \includegraphics[width=0.3\textwidth]{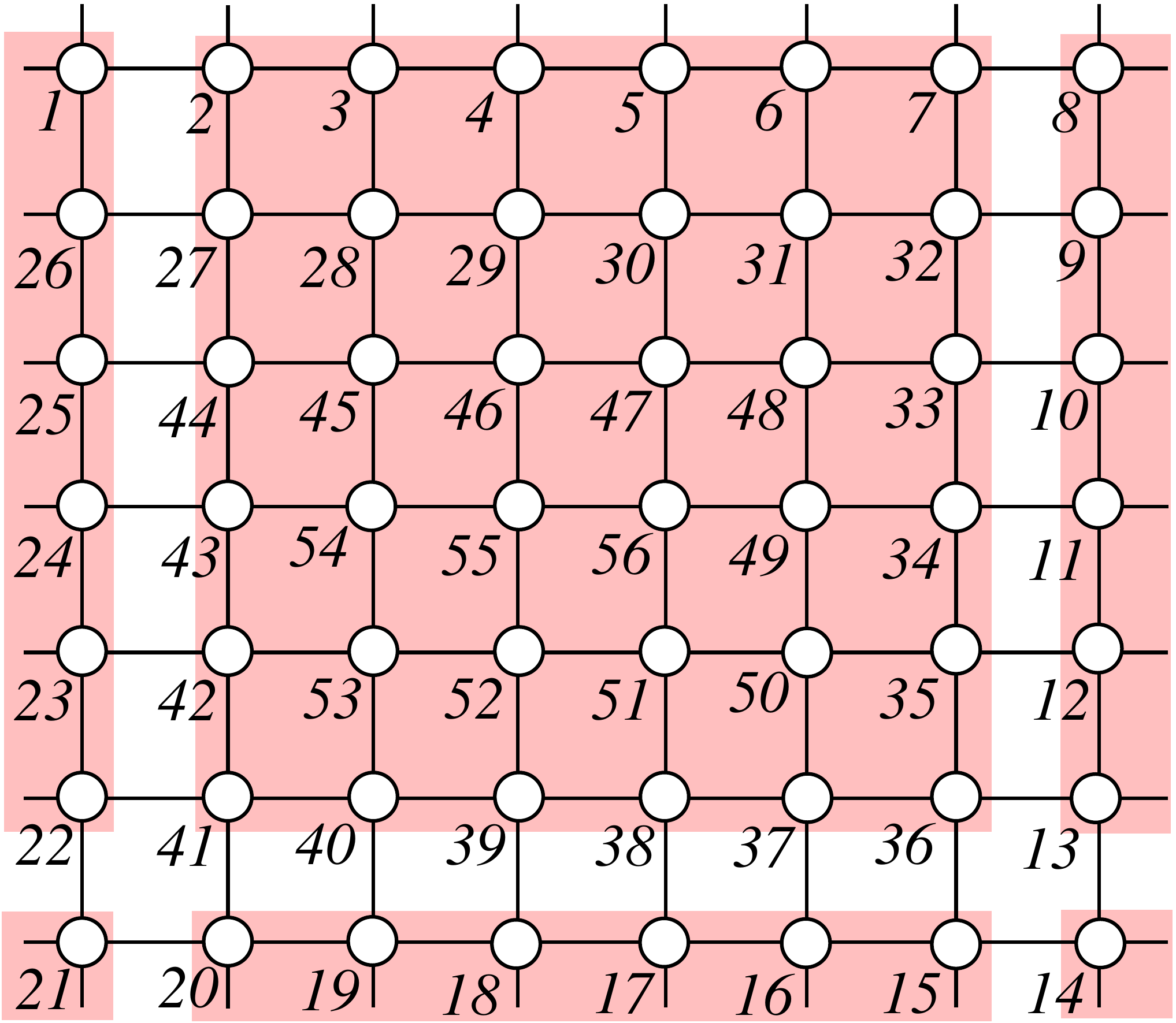}
    \end{center}
    \caption{
      \label{fig:msupdaten6}
      The central square region contains $n\times n$ vertices ($n=6$) and it
      interacts with the three neighboring square regions (partly
      shown) on the left, bottom, and right side.
    }
  \end{figure}

  To perform region-graph BP or loop-corrected BP iteration on a square
  lattce, the most demanding task is computing the joint probability
  distribution of spin states for the vertices on the boundary of a region.
  Let us consider the concrete example shown in Fig.~\ref{fig:msupdaten6}.
  The central (C) square region contains $n\times n$ vertices with $n=6$, and
  it receives messages from three other square regions on the left
  (L), bottom (B), and right (R) side. Denote by
  $\underline{\sigma}_{T} \equiv (\sigma_2, \sigma_3, \ldots, \sigma_{7})$ a
  generic spin configuration for the $n$ vertices on the top (T) boundary of
  the central region. This spin configuration is affected by the interactions
  within the central region and the interactions between the central region and
  the three neighboring regions, and its probability distribution
  $P_T(\underline{\sigma}_T)$ is expressed as
  \begin{eqnarray}
    & & \hspace*{-0.85cm} P_{T}(\underline{\sigma}_T) \propto 
    \sum\limits_{\underline{\sigma}_{C\backslash T}}
    e^{-\beta E_C} \Bigl[
      \sum\limits_{\underline{\sigma}_L}
      P_{L}(\underline{\sigma}_L)
      e^{-\beta E_{L,C}} \Bigr] \nonumber \\
    & & \hspace*{-0.2cm} \times \Bigl[
      \sum\limits_{\underline{\sigma}_B}
      P_{B}(\underline{\sigma}_B)
    e^{-\beta E_{B,C}} \Bigr]
    \Bigl[
      \sum\limits_{\underline{\sigma}_{R}}
      P_{R}(\underline{\sigma}_R)
      e^{-\beta E_{R,C}} \Bigr] \; . 
    \label{eq:regionsum}
  \end{eqnarray}
  In this expression, 
  $\underline{\sigma}_{C\backslash T} \equiv (\sigma_{27}, \sigma_{28}, \ldots,
  \sigma_{55}, \sigma_{56})$
  is a spin configuration for all the other $(n-1)\times n$ vertices of the
  central region except the $n$ vertices at the top boundary, and $E_C$ is the
  total internal energy of this central region;
  $\underline{\sigma}_{L} \equiv (\sigma_1, \sigma_{26}, \ldots, \sigma_{22})$
  is a spin configuration for the $n$ boundary vertices of the left
  region, and $P_L(\underline{\sigma}_L)$ is an input probability
  distribution of $\underline{\sigma}_L$, and $E_{L, C}$ is the interation
  energy between the left and the central region; similarly,
  $\underline{\sigma}_{R} \equiv (\sigma_{13}, \sigma_{12}, \ldots, \sigma_{8})$
  is a spin configuration for the boundary vertices of the right region,
  $P_R(\underline{\sigma}_R)$ is an input probability
  distribution of $\underline{\sigma}_R$, $E_{R, C}$ is the interaction energy
  between the right and the central region, and 
  $\underline{\sigma}_{B} \equiv (\sigma_{20}, \sigma_{19}, \ldots, \sigma_{15})$
  is a spin configuration for the boundary vertices of the bottom
  region, $P_B(\underline{\sigma}_B)$ is an input probability
  distribution of $\underline{\sigma}_B$, $E_{B, C}$ is the interaction energy
  between the bottom and the central region.
  Notice that the LC-BP equations (\ref{eq:lcBPmax2}), (\ref{eq:lcBPmax2c}),
  and (\ref{eq:lcBPregion}) all have the same form of Eq.~(\ref{eq:regionsum}).

  %figure 07

  According to Eq.~(\ref{eq:regionsum}), one needs to sum over a total number
  of $2^{n(n+2)}$ different spin configurations to determine the output
  probability $P_T(\underline{\sigma}_T)$ of a single spin configuration
  $\underline{\sigma}_T$. A naive application of Eq.~(\ref{eq:regionsum}) is
  therefore feasible only for very small values of $n$ (e.g., $n\leq 3$).
  
  We now introduce a numerical trick that greatly accelerate this summation
  process. By this simple trick we reduce the total number of needed
  operations to sum over all the spin configurations from $O(2^{n(n+2)})$
  to $O( n^2 2^{n} )$, and also dramatically reduce the total amount of
  storage space needed in the numerical computation.

  First we notice that, due to the binary nature of the spins, a generic
  probability distribution $p(\sigma_1, \sigma_2, \ldots, \sigma_n)$ over
  $n$ spins can be written in the following form:
  \begin{equation}
    \label{eq:binaryP}
    p(\sigma_1,  \ldots, \sigma_n) =
    \sum\limits_{s_1=0}^{1} \sum\limits_{s_2=0}^{1}
    \ldots \sum\limits_{s_n=0}^{1}
    c_{s_1 s_2 \ldots s_n} \sigma_1^{s_1} \sigma_2^{s_2} \ldots \sigma_n^{s_n}
    \; ,
  \end{equation}
  where $s_i\in \{0, 1\}$ for $i=1,2,\ldots, n$ and
  $\{c_{s_1 s_2 \ldots s_n}\}$ is a set of $2^{n}$ coefficients, with
  $c_{0 0 \ldots 0} \equiv 2^{- n}$ due to the normalization constraint. 
  Therefore the probability distribution
  $p(\sigma_1, \sigma_2, \ldots, \sigma_n)$ is completely characterized by
  the coefficient set  $\{c_{s_1 s_2 \ldots s_n}\}$.

  Due to the fact that 
  \begin{subequations}
    \begin{align}
      e^{\beta h_i \sigma_i} & \equiv 
      \cosh (\beta h_i ) \bigl[ 1 + \tanh (\beta h_i) \sigma_i \bigr] \; , \\
      e^{\beta J_{i i^\prime} \sigma_i \sigma_{i^\prime}} & \equiv
      \cosh (\beta J_{i i^\prime}) \bigl[ 1  + \tanh (\beta J_{i i^\prime}) \sigma_i
        \sigma_{i^\prime} \bigr]  \; ,
    \end{align}
  \end{subequations}
  then for $i, j\in \{1, 2,\ldots, n\}$ ($i<j$)
  and $i^\prime \notin
  \{1, 2, \ldots, n\}$,
  \begin{subequations}
    \label{eq:changecset}
    \begin{align}
      &  e^{\beta h_i \sigma_i} p(\sigma_1, \ldots, \sigma_n) =
      \cosh(\beta h_i) \sum\limits_{s_1 s_2 \ldots s_n} 
      \sigma_1^{s_1} \sigma_2^{s_2} \ldots s_n^{s_n} \nonumber \\
      & \quad \times 
      \bigl[c_{s_1 s_2 \ldots s_n}+\tanh(\beta h_i) c_{s_1 \ldots s_{i-1}
          \overline{s}_{i} s_{i+1} \ldots s_n} \bigr]
      \; , \label{eq:changecseta}
      \\
      & \sum{_{\sigma_i}}
      e^{\beta J_{i i^\prime} \sigma_i \sigma_{i^\prime}} p(\sigma_1, \ldots, \sigma_n) =
      2  \cosh(\beta J_{i i^\prime}) \nonumber \\
      & \quad \times \sum\limits_{s_1 s_2 \ldots s_n} 
      \sigma_1^{s_1} \ldots \sigma_{i-1}^{s_{i-1}} \sigma_{i^\prime}^{s_i} 
      \sigma_{i+1}^{s_{i+1}} \ldots s_n^{s_n} \nonumber \\
      & \quad \quad \times 
      \bigl[(1-s_i + s_i \tanh(\beta J_{i i^\prime}) \bigr] c_{s_1 s_2 \ldots s_n}
      \; , \label{eq:changecsetb} \\
      &  e^{\beta J_{i j} \sigma_i \sigma_j} p(\sigma_1, \ldots, \sigma_n) =
      \cosh(\beta J_{i j}) \times \nonumber \\
      & \quad \sum\limits_{s_1 s_2 \ldots s_n} 
      \sigma_1^{s_1} \sigma_2^{s_2} \ldots s_n^{s_n} 
      \bigl[c_{s_1 s_2 \ldots s_n}+ \nonumber \\
        & \quad \quad \quad \tanh(\beta J_{i j})
        c_{s_1 \ldots s_{i-1} \overline{s}_i s_{i+1} \ldots s_{j-1}
          \overline{s}_j s_{j+1} \ldots s_n}\bigr]
      \; , \label{eq:changecsetc}
    \end{align}
  \end{subequations}
  where $\overline{s}_i =1$ if $s_i=0$ and $\overline{s}_i = 0$ if $s_i=1$.
  Equation (\ref{eq:changecset}) therefore gives a set of rules on how the
  coefficients set $\{c_{s_1 s_2 \ldots s_n}\}$ changes as
  $p(\sigma_1, \ldots, \sigma_n)$ is perturbed by multiplication and summation.

\begin{figure}
    \begin{center}
      \includegraphics[width=0.2\textwidth]{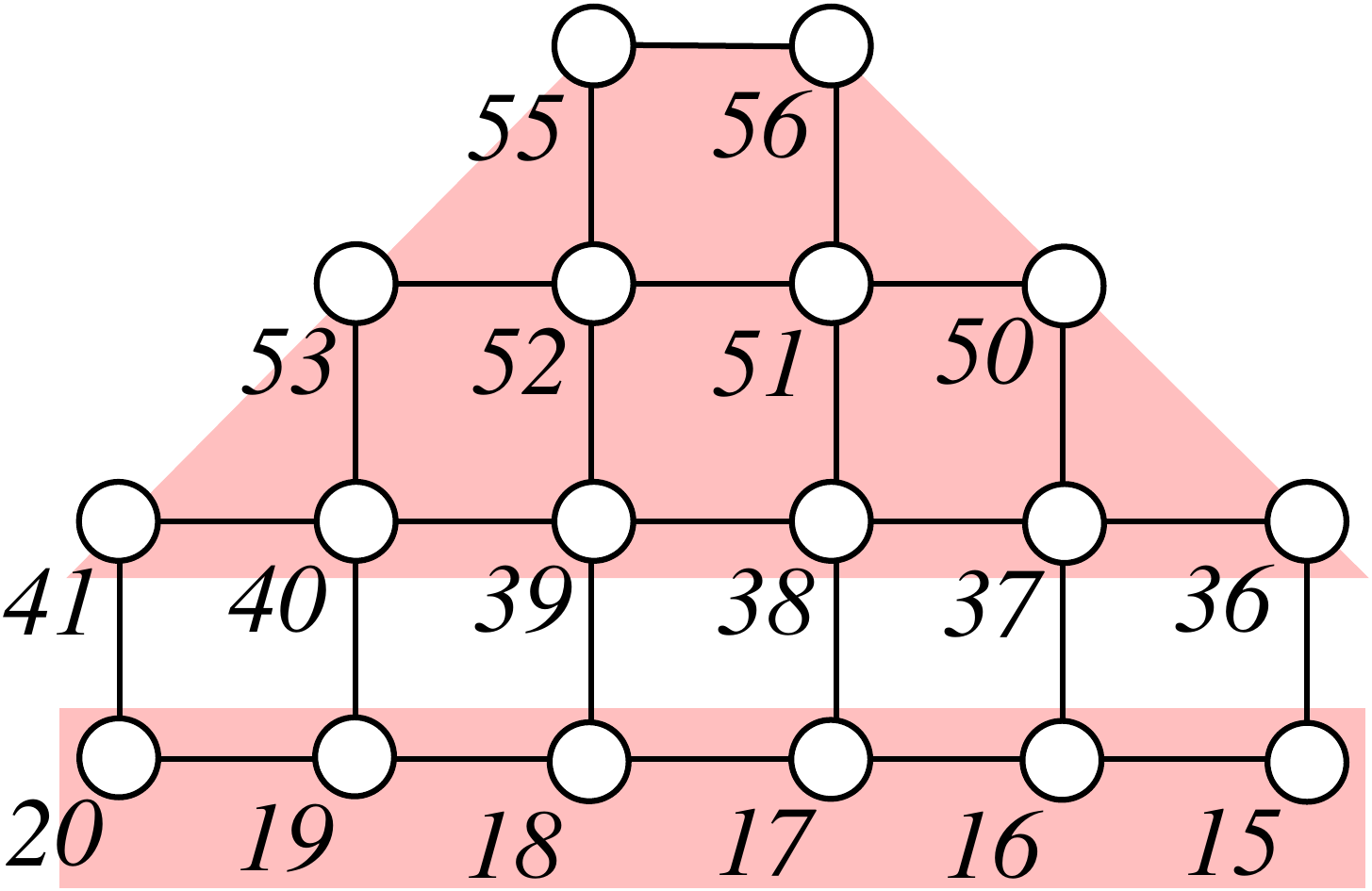}
    \end{center}
    \caption{
      \label{fig:inputBottom}
      Given an input probability distribution 
      $P_B(\sigma_{20}, \ldots, \sigma_{15})$ for the set
      $\{20, 19, \ldots, 15\}$ of vertices on the bottom row, the
      probability distribution $Q_B(\sigma_{41}, \ldots, \sigma_{36})$
      for the set $\{41, 53, 55, 56, 50, 36\}$ of boundary vertices 
      can be determined  by recursion from the bottom row up to the top row.
    }
  \end{figure}

  We simplify the computation of Eq.~(\ref{eq:regionsum}) by treating the
  three input probability distributions separately. 
  For example, starting from the input probability distribution
  $P_B(\sigma_{20}, \sigma_{19}, \ldots, \sigma_{15})$ of the bottom region
  (see Fig.~\ref{fig:inputBottom}), we obtain a probability
  distribution $Q_B(\sigma_{41}, \sigma_{53}, \ldots, \sigma_{36})$
  for the set of $n$ boundary vertices $\{41, 53, 55, 56, 50, 36\}$
  through the following recursive process:
  (1) initialize the coefficients set of $Q_B(\cdot)$ to be identical to that
  of $P_B(\cdot)$;
  (2) then consider sequentially all the $n$ vertical edges
  $\langle 20, 41\rangle$, $\langle 19, 40\rangle$, ...,
  $\langle 15, 36 \rangle$ between the central and the bottom region
  and modify the coefficients set of $Q_B(\cdot)$ according to
  Eq.~(\ref{eq:changecsetb});
  (3) then consider sequentially all the $(n-1)$ horizontal edges
  $\langle 41, 40\rangle$, $\langle 40, 39 \rangle$, ...,
  $\langle 37, 36\rangle$ between the set of vertices
  $\{41, 40, 39, 38, 37, 36\}$ and further modify the coefficients
  set of $Q_B(\cdot)$ according to Eq.~(\ref{eq:changecsetc});
  (4) then consider sequentially all the $(n-1)$ external fields on the
  set of internal vertices $\{40, 39, \ldots, 37\}$ and further modify the 
  coefficients set of $Q_B(\cdot)$ according to Eq.~(\ref{eq:changecseta});
  (5) repeat the previous three steps on the row containing the set of
  vertices $\{53, 52, 51, 50\}$: apply Eq.~(\ref{eq:changecsetb}) on
  the set of vertical edges 
  $\{\langle 40, 53\rangle, \ldots, \langle 37, 50\rangle\}$ and then
  apply Eq.~(\ref{eq:changecsetc}) on the  horizontal edges
  $\langle 53, 52\rangle$, $\langle 52, 51\rangle$ and 
  $\langle 51, 50\rangle\}$, and then apply Eq.~(\ref{eq:changecseta}) on
  the internal vertices $52$ and $51$;
  (6) finally, apply Eq.~(\ref{eq:changecsetb}) on the edges
  $\langle 52, 55\rangle$ and $\langle 51, 56\rangle$, and apply
  Eq.~(\ref{eq:changecsetc}) on edge $\langle 55, 56\rangle$
  and then output the coefficients set of
  $Q_B(\sigma_{41},\sigma_{53},\sigma_{55},\sigma_{56},\sigma_{50},\sigma_{36})$.
  
  % figure 08
  
  The joint probability distributions $Q_L(\sigma_{2}, \ldots, \sigma_{41})$
  for the set of vertices $\{2, 28, 46, 55, 53, 41\}$ and 
  $Q_R(\sigma_{36}, \ldots, \sigma_{7})$ for the set of vertices
  $\{36, 50, 56, 47, 31, 7\}$, see Fig.~\ref{fig:msupdaten6},  are obtained
  through the same recursive process starting from $P_L(\cdot)$ and
  $P_R(\cdot)$, respectively. The only additional feature is that we now need
  to consider the external fields of all the vertices in these two boundary
  sets (again through applying Eq.~(\ref{eq:changecseta}) to $Q_L(\cdot)$ and
  $Q_R(\cdot)$ repeatedly).
  
  With these preparations, we then obtain a joint probability distribution
  $Q(\sigma_{2}, \ldots, \sigma_{7})$ for the set of vertices
  $\{2, 28, 46, 47, 31, 7\}$ through the following expression:
  \begin{eqnarray}
    & &  \hspace*{-0.5cm} Q(\sigma_{2}, \sigma_{28}, \sigma_{46}, \sigma_{47},
    \sigma_{31}, \sigma_7)  \propto
    e^{\beta J_{46, 47} \sigma_{46} \sigma_{47}} \times \nonumber \\
    & & \sum\limits_{\sigma_{41}, \sigma_{53}, \sigma_{55}}
    \sum\limits_{\sigma_{56}, \sigma_{50}, \sigma_{36}}
    Q_L(\sigma_2, \sigma_{28}, \sigma_{46},
    \sigma_{55}, \sigma_{53}, \sigma_{41}) \nonumber \\
    & & \quad \quad \times 
    Q_B(\sigma_{41}, \sigma_{53}, \sigma_{55},
    \sigma_{56}, \sigma_{50}, \sigma_{36}) \nonumber \\
    & & \quad \quad \times
    Q_R(\sigma_{36}, \sigma_{50}, \sigma_{56},
    \sigma_{47}, \sigma_{31}, \sigma_{7}) \nonumber \\
    & & \propto 
    e^{\beta J_{46, 47} \sigma_{46} \sigma_{47}}
    \sum\limits_{s_{2} s_{28} s_{46} s_{47} s_{31} s_7} 
    \sigma_{2}^{s_{2}} \sigma_{28}^{s_{28}} \sigma_{46}^{s_{46}}
    \sigma_{47}^{s_{47}} \sigma_{31}^{s_{31}} \sigma_{7}^{s_{7}}
    \nonumber \\
    & & \quad \quad \times \sum\limits_{s_{55} s_{53} s_{41}}
    \sum\limits_{s_{36} s_{50} s_{56}}
    c_{s_{2} s_{28} s_{46} s_{55} s_{53} s_{41}}^{(L)} \nonumber \\
    & & \quad \quad \quad \quad  \times
    c_{s_{41} s_{53} s_{55} s_{56} s_{50} s_{36}}^{(B)}
    c_{s_{36} s_{50} s_{56} s_{47} s_{31} s_{7}}^{(R)} \; .
    \label{eq:Qexp}
  \end{eqnarray}
  In the above expression, the coefficient sets
  $\{c_{s_2 \cdots s_{41}}^{(L)}\}$, $\{c_{s_{41} \cdots s_{36}}^{(B)}\}$, and
 $\{c_{s_{36} \cdots s_{7}}^{(R)}\}$ correspond to $Q_L(\cdot)$,
  $Q_B(\cdot)$, and $Q_R(\cdot)$, respectively.
  The effect of the multiplication term $e^{\beta J_{46, 47} \sigma_{46} \sigma_{47}}$
  to the coefficient set of the probability distribution $Q(\cdot)$
  can again be obtained through Eq.~(\ref{eq:changecsetc}).

\begin{figure}
    \begin{center}
      \includegraphics[width=0.2\textwidth]{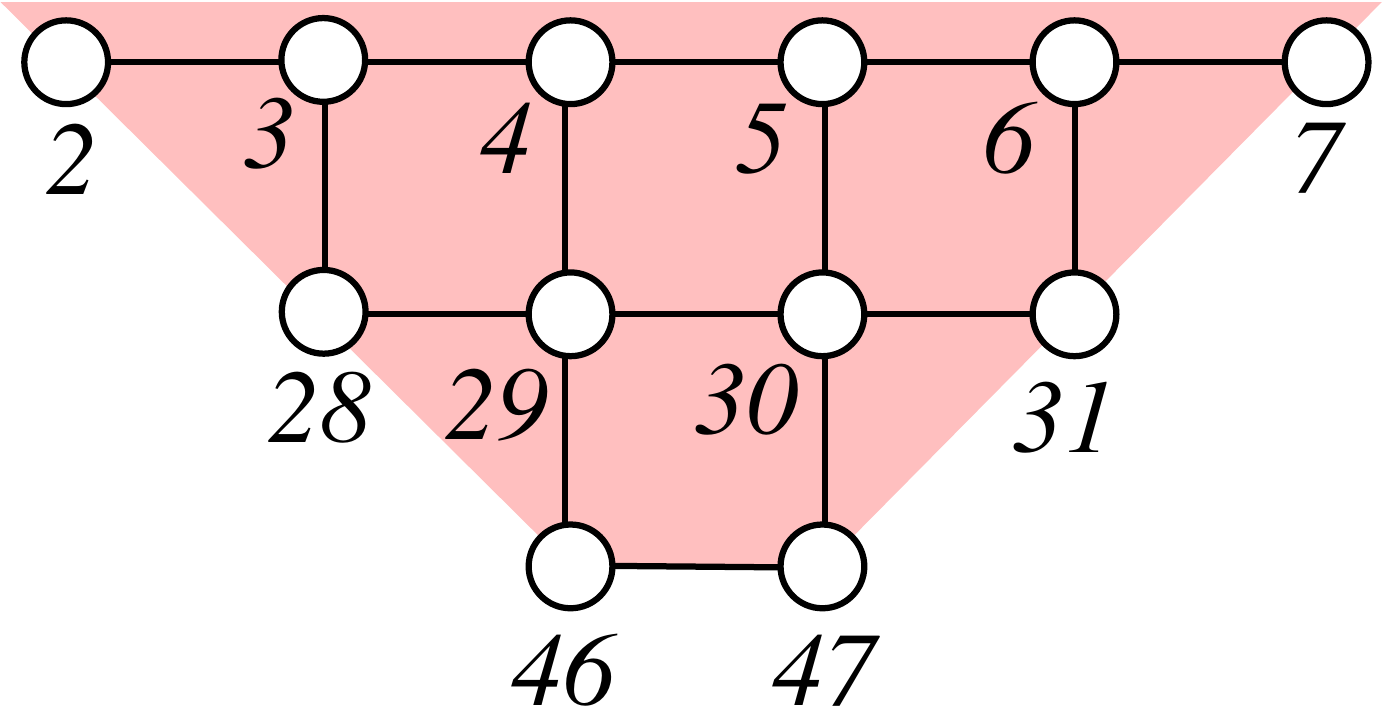}
    \end{center}
    \caption{
      \label{fig:sumup}
      Given an input joint probability distribution 
      $Q(\sigma_{2}, \ldots, \sigma_{7})$ for the set
      $\{2, 28, 46, 47, 31, 7\}$ of vertices on the bottom boundary,
      the joint probability distribution $P_T(\sigma_{2}, \ldots, \sigma_{7})$
      for the set $\{2, 3, 4, 5, 6, 7\}$ on the top row
      can be determined recursively from the bottom row up to the top row.
    }
  \end{figure}

  Finally, the  probability distribution 
  $P_T(\sigma_2, \ldots, \sigma_7)$ for the set $\{2, 3, 4, 5, 6, 7\}$ of
  vertices at the top boundary is determined from
  $Q(\sigma_2, \sigma_{28}, \sigma_{46}, \sigma_{47}, \sigma_{31}, \sigma_7)$
  through the following recursive process  (see Fig.~\ref{fig:sumup}):
  (1) set the coefficients set of $P_T(\cdot)$ to be identical to that of
  $Q(\cdot)$;
  (2) then consider the vertical edges $\langle 29, 46\rangle$
  and $\langle 30, 47\rangle$ sequentially and modify the coefficients set of
  $P_T(\cdot)$ according to Eq.~(\ref{eq:changecsetb});
  (3) then consider all the horizontal edges $\langle 28, 29\rangle$,
  $\langle 29, 30 \rangle$, and $\langle 30, 31\rangle$ between the
  set of vertices $\{28, 29, 30, 31\}$ and the external
  fields on vertices $29$ and $30$  and further modify the coefficients
  set of $P_T(\cdot)$ according to Eq.~(\ref{eq:changecsetc}) and
  Eq.~(\ref{eq:changecseta}), respectively; (4) repeat the 
  operations of steps (2) and (3) on the vertical edges between the top and
  the second row of Fig.~\ref{fig:sumup}, the horizontal edges of the top
  row, and the set of vertices $\{3,4,5,6\}$. We then output the resulting
  coefficient set of $P_T(\cdot)$ as the result of original computing
  task Eq.~(\ref{eq:regionsum}).
  
  %figure 09

It is straightforward to extend the numerical trick of this appendix to other
values of even $n$ and also to the cases of $n$ being odd.
For studying lattice models on a three-dimensional cubic lattice, this same
trick can be applied to a cubic region containing $n\times n \times n$
vertices.

\section*{Appendix B: Loop-corrected belief propagation with memory 
  capacity $C=3$}

\begin{figure}
  \begin{center}
    \includegraphics[width=0.475\textwidth]{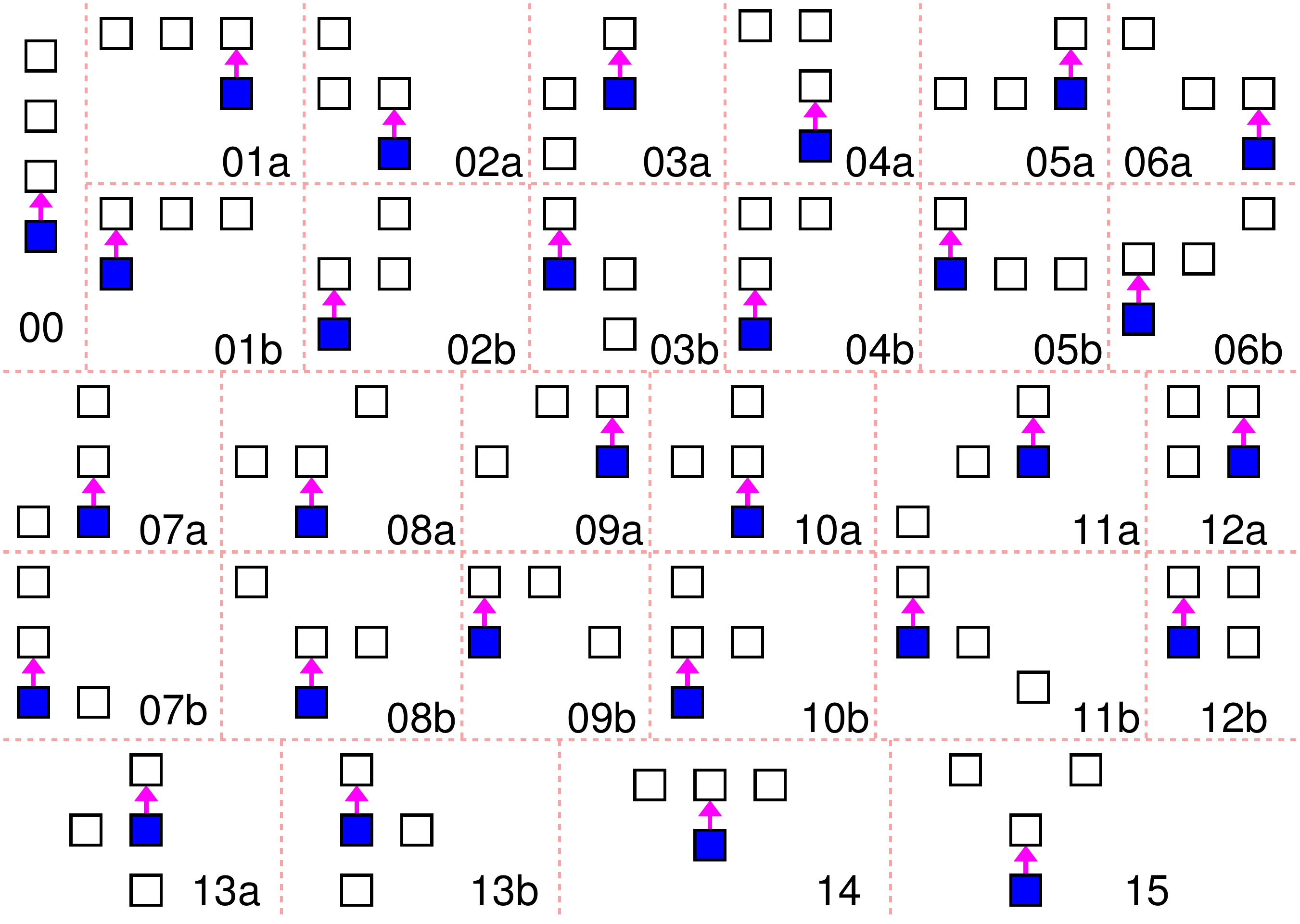}
  \end{center}
  \caption{
    \label{fig:LCBPm3}
    When the memory capacity is set to $C=3$, each focal vertex/region
    (denoted by a filled small square) needs to remember the positions of
    the other three deleted vertices/regions 
    (denoted by three unfilled small squares). In total we need to 
    distinguish $29$ different patterns of the three deleted 
    vertices/regions, which are indexed as $00$, $01a$ and $01b$, $\ldots$,
    $14$, and $15$. The small arrows indicate the cavity message of the
    focal vertex/region to the deleted vertices/regions.
    For the purpose of clarity we separate different patterns through
    the thin dashed lines.
  }
\end{figure}

When the memory capacity is set to $C=3$, then with respective to a focal
vertex or region (denoted by a filled small square in each block of
Fig.~\ref{fig:LCBPm3}), we need to consider $29$ different patterns of
the three deleted vertices or regions (denoted by three unfilled small 
squares in each block of Fig.~\ref{fig:LCBPm3}). 
These $29$ patterns are indexed as $00$, $01a$ and $01b$, $02a$ and 
$02b$, $\ldots$, $13a$ and $13b$, $14$, and $15$ in
Fig.~\ref{fig:LCBPm3} for the convenience of discussion. The patterns $01a$ 
and $01b$ (and similarly $02a$ and $02b$, $03a$ and $03b$, ...)
are related by a mirror symmetry. 

%figure 10

Each pattern of Fig.~\ref{fig:LCBPm3} is associated with a cavity message.
For example,  suppose
vertices $l$, $m$, $n$ are deleted from the graph $G$ of
Fig.~\ref{fig:2Dsquare}, then the pattern $01a$ of Fig.~\ref{fig:LCBPm3}
corresponds to the cavity message
$q_{s\backslash\{l,m,n\}}(\sigma_s)$ from vertex $s$ to vertex $n$, while
pattern $01b$ corresponds to the cavity message
$q_{q\backslash\{l,m,n\}}(\sigma_q)$ from vertex $q$ to vertex $l$.
As another example at the region graph level, 
suppose regions $\gamma_2$, $\gamma_6$ and
$\gamma_9$ are deleted from the region graph $\mathcal{R}$ of
Fig.~\ref{fig:2Dregion}, then the pattern $03b$ of Fig.~\ref{fig:LCBPm3}
corresponds to the cavity message
$q_{\gamma_5\backslash \{\gamma_2, \gamma_6, \gamma_9\}}(\sigma_m,\sigma_n)$
from region $\gamma_5$ to $\gamma_2$.

% figure 11
\begin{figure}
  \begin{center}
    \includegraphics[width=0.475\textwidth]{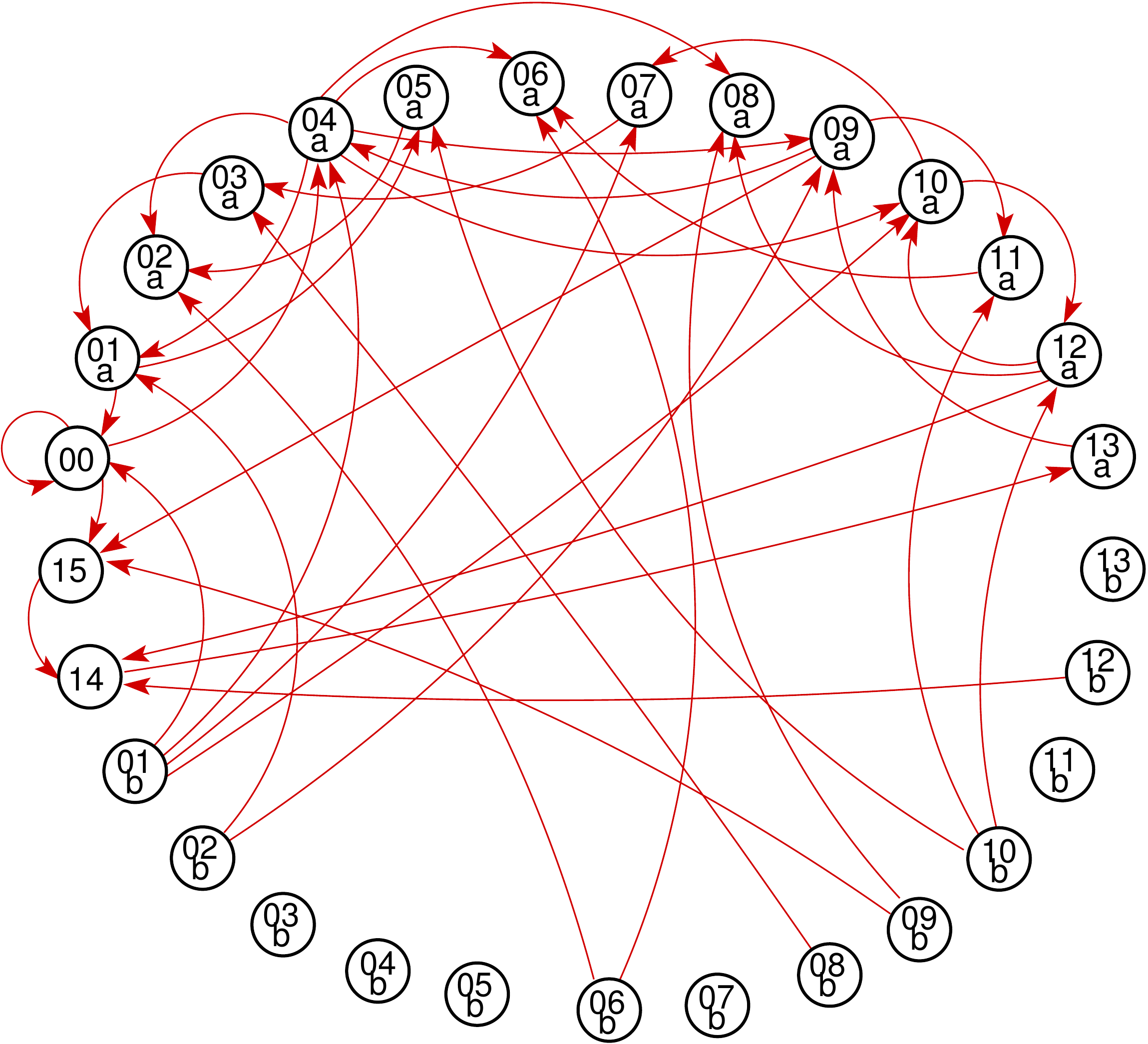}
  \end{center}
  \caption{
    \label{fig:LCBPm3network}  
    Diagram showing how the cavity message of all the $29$ patterns in
    Fig.~\ref{fig:LCBPm3} are iteratively determined (see main text for
    more details). For reason of clarity, for each pair of mirror patterns 
    (say $01a$ and $01b$) we only draw the input edges to one of the
    patterns ($01a$) but not to the mirror pattern ($01b$). The edges to
    each mirror pattern can be easily constructed by symmetry 
    considerations. For example, since pattern $01a$ receives edges
    from patterns $02b$,
    $03a$ and $04a$, then pattern $01b$ must receive edges from
    patterns $02a$, $03b$ and $04b$.}
\end{figure}

The iteration of the $29$ cavity messages for the $29$ patterns of
Fig.~\ref{fig:LCBPm3} is carried out following the updating diagram of
Fig.~\ref{fig:LCBPm3network}. Each directed edge $p_1 \rightarrow p_2$
in this diagram points from one pattern (say $p_1=04a$) to another 
pattern (say $p_2=01a$), and it means that the cavity message of
pattern $p_2$ is determined (partly) from the cavity message of pattern
$p_1$. For example, there are three directed edges
(from patterns $01a$, $01b$ and $00$, respectively) to pattern $00$, 
meaning that the output cavity message of pattern $00$ can be computed
based on three inputing cavity messages from patterns $00$, $01a$ and
$01b$. In the specific case of Fig.~\ref{fig:2Dsquare},
we have
\begin{eqnarray}
  & &  q_{r\backslash \{c,h,m\}}(\sigma_r) \propto
  e^{\beta h^0_r \sigma_r} \bigl[\sum{_{\sigma_q}}
    e^{\beta J_{q r} \sigma_q \sigma_r}
    q_{q\backslash \{h,m,r\}}    (\sigma_q) 
    \bigr]
  \nonumber \\
  & & \quad \quad \quad \quad \times
  \bigl[\sum{_{\sigma_w}} e^{\beta J_{w r} \sigma_w \sigma_r} 
    q_{w\backslash \{h,m,r\}}(\sigma_w) \bigr]
  \nonumber \\
  & & \quad \quad \quad \quad \times
  \bigl[\sum{_{\sigma_s}}    e^{\beta J_{s r} \sigma_s \sigma_r}
    q_{s\backslash \{h,m,r\}}(\sigma_s) \bigr] \; .
\end{eqnarray}
The updating equations for the other $28$ cavity messages can be written
down in a similar way according to Fig.~\ref{fig:LCBPm3network}. Notice
that in Fig.~\ref{fig:LCBPm3network} we only draw the input edges to
patterns
$00$, $14$, $15$ and patterns $01a$, $02a$, $\ldots$, $13a$ but not the
input edges to all the mirror patterns $01b$, $02b$, $\ldots$, $13b$
to avoid the diagram being too complicated. We can easily construct
all the missing directed edges by symmetry considerations. For example,
since pattern $03a$ receives edges from patterns $07a$ and $08b$,
then pattern $03b$ must receive edges from patterns $07b$ and $08a$.
In the specific case of regions 
$\gamma_2$, $\gamma_6$, and $\gamma_9$ being deleted from 
Fig.~\ref{fig:2Dregion}, we have
\begin{eqnarray}
  & &  q_{\gamma_5\backslash \{\gamma_2, \gamma_6, \gamma_9\}}(\sigma_m,
  \sigma_n) \propto
  e^{\beta h^0_m \sigma_m + \beta h^0_n \sigma_n + \beta J_{m n} \sigma_m \sigma_n}
\nonumber \\
& & \quad  \times 
\sum\limits_{\sigma_r, \sigma_s}
\bigl[\sum\limits_{\sigma_l, \sigma_q} e^{\beta J_{l m} \sigma_l \sigma_m +
    \beta J_{q r} \sigma_q \sigma_r} q_{\gamma_4 \backslash 
    \{\gamma_2, \gamma_5, \gamma_9\}}
  (\sigma_l, \sigma_q) \bigr]
\nonumber \\
& & \quad \quad \quad \times
\bigl[\sum\limits_{\sigma_w, \sigma_x} e^{\beta J_{w r} \sigma_w \sigma_r +
    \beta J_{x s} \sigma_x \sigma_s} q_{\gamma_8 \backslash 
    \{\gamma_2, \gamma_5, \gamma_9\}}
  (\sigma_w, \sigma_x) \bigr] \; . \nonumber \\
& & 
\end{eqnarray}

\end{appendix}

%\bibliography{/Users/zhouhj/references.bib}

\end{document}